\documentclass[iop,apj]{emulateapj}
\usepackage{apjfonts}
\usepackage{epsf}
\newcommand{\Msun}{\mathrm{M}_{\odot}}

\newcommand{\mav}{\overline{m}}
\newcommand{\logMav}{\overline{\log{M}}}
\newcommand{\feh}{\mathrm{[Fe/H]}}

\begin{document}

\slugcomment{Accepted to ApJ}
\shortauthors{Muratov and Gnedin}
\shorttitle{Metallicity Distribution of Globular Clusters}

\title{Modeling the Metallicity Distribution of Globular Clusters}
        
\author{Alexander L. Muratov and Oleg Y. Gnedin}

\affil{University of Michigan, Department of Astronomy, Ann Arbor, MI 48109\\
    \mbox{\tt muratov@umich.edu, ognedin@umich.edu}}

\date{\today}

\begin{abstract}
Observed metallicities of globular clusters reflect physical
conditions in the interstellar medium of their high-redshift host
galaxies.  Globular cluster systems in most large galaxies display
bimodal color and metallicity distributions, which are often
interpreted as indicating two distinct modes of cluster formation.
The metal-rich and metal-poor clusters have systematically different
locations and kinematics in their host galaxies.  However, the red and
blue clusters have similar internal properties, such as the masses,
sizes, and ages.  It is therefore interesting to explore whether both
metal-rich and metal-poor clusters could form by a common mechanism
and still be consistent with the bimodal distribution.  We present
such a model, which prescribes the formation of globular clusters
semi-analytically using galaxy assembly history from cosmological
simulations coupled with observed scaling relations for the amount and
metallicity of cold gas available for star formation.  We assume that
massive star clusters form only during mergers of massive gas-rich
galaxies and tune the model parameters to reproduce the observed
distribution in the Galaxy.  A wide, but not entire, range of model
realizations produces metallicity distributions consistent with the
data.  We find that early mergers of smaller hosts create exclusively
blue clusters, whereas subsequent mergers of more massive galaxies
create both red and blue clusters.  Thus bimodality arises naturally
as the result of a small number of late massive merger events.  This
conclusion is not significantly affected by the large uncertainties in
our knowledge of the stellar mass and cold gas mass in high-redshift
galaxies.  The fraction of galactic stellar mass locked in globular
clusters declines from over 10\% at $z>3$ to 0.1\% at present.
\end{abstract}


\section{Introduction}

A self-consistent description of the formation of globular clusters
remains a challenge to theorists.  A particularly puzzling observation
is the apparent bimodality, or even multimodality, of the color
distribution of globular cluster systems in galaxies ranging from
dwarf disks to giant ellipticals \citep[reviewed
by][]{brodie_strader06}.  This color bimodality likely translates into
a bimodal distribution of the abundances of heavy elements such as
iron.  We know this to be the case in the Galaxy as well as in M31,
where relatively accurate spectral measurements exist for a large
fraction of the clusters.  In this paper we will interchangeably refer
to metal-poor clusters as blue clusters, and to metal-rich clusters as
red clusters.

Bimodality in the globular cluster metallicity distribution of
luminous elliptical galaxies was proposed by \citet{zepf_ashman93},
following a theoretical model of \citet{ashman_zepf92}.  The concept
of cluster bimodality became universally accepted because the two
populations also differ in other observed characteristics.  The system
of red clusters has a significant rotation velocity similar to the
disk stars whereas blue clusters have little rotational support, in
the three disk galaxies observed in detail: Milky Way, M31, and M33
\citep{zinn85}.  In elliptical galaxies, blue clusters have a higher
velocity dispersion than red clusters, both due to lack of rotation
and more extended spatial distribution.  Red clusters are usually more
spatially concentrated than blue clusters \citep{brodie_strader06}.
All of these differences, however, are in external properties
(location and kinematics), which reflect {\it where} the clusters
formed, but not {\it how}.  The internal properties of the red and
blue clusters are similar: masses, sizes, and ages, with only slight
differences.  Even the metallicities themselves differ typically by a
factor of 10 between the two modes, not enough to affect the dynamics
of molecular clouds from which these clusters formed.  Could it be
then that both red and blue clusters form in a similar way on small
scales, such as in giant molecular clouds, while the differences in
their metallicity and spatial distribution reflect when and where such
clouds assemble?

All scenarios proposed in the literature assumed different formation
mechanisms for the red and blue clusters, and most scenarios
envisioned the stellar population of one mode to be tightly linked to
that of the host galaxy \citep[e.g.,][]{forbes_etal97, cote_etal98,
strader_etal05, griffen_etal10}.  The other mode is assumed to have
formed differently, in some unspecified ``primordial'' way.  This
assumption only pushed the problem back in time but it did not solve
it.  For example, \citet{beasley_etal02} used a semi-analytical model
of galaxy formation to study bimodality in luminous elliptical
galaxies and needed two separate prescriptions for the blue and red
clusters.  In their model, red clusters formed in gas-rich mergers
with a fixed efficiency of 0.007 relative to field stars, while blue
clusters formed in quiescent disks with a different efficiency of
0.002.  The formation of blue clusters also had to be artificially
truncated at $z=5$.  \citet{strader_etal05}, \citet{rhode_etal05}, and
\citet{griffen_etal10} suggested that the blue clusters could instead
have formed in very small halos at $z \ga 10$, before cosmic
reionization removed cold gas from such halos.  This scenario requires
high efficiency of cluster formation in the small halos and also
places stringent constraints on the age spread of blue clusters to be
less than 0.5 Gyr.  Unfortunately, even the most recent measurements
of relative cluster ages in the Galaxy \citep{deangeli_etal05,
marin-franch_etal09, dotter_etal10} cannot detect age differences
smaller than 9\%, or about 1 Gyr, and therefore cannot support or
falsify the reionization scenario.  \citet{dotter_etal10} also show
that the red clusters have larger dispersion of ages (15\%, or about 2
Gyr) and those located outside 15 kpc of the Galactic center tend to
show measurably lower ages, by as much as 50\% (or 6 Gyr).  In
addition, \citet{strader_etal09} find that the red clusters in M31
have lower mass-to-light ratios than the blue clusters, possibly
indicating an age variation.

In this paper we set out to test whether a common mechanism could
explain the formation of both modes and produce an entire metallicity
distribution consistent with the observations.  We begin with a
premise of the hierarchical galaxy formation in a $\Lambda$CDM
universe.  Hubble Space Telescope observations have convincingly
demonstrated one of the likely formation routes for massive star
clusters today -- in the mergers of gas-rich galaxies
\citep[e.g.,][]{holtzman_etal92, oconnell_etal95, whitmore_etal99,
zepf_etal99}.  We adopt this single formation mechanism for our model
and assume that clusters form only during massive gas-rich mergers.
We follow the merging process of progenitor galaxies in a Galaxy-sized
environment using a set of cosmological $N$-body simulations.  We need
to specify what type and how many clusters will form in each merger
event.  For this purpose, we use observed scaling relations to assign
each dark matter halo a certain amount of cold gas that will be
available for star formation throughout cosmic time and an average
metallicity of that gas.  In order to keep the model transparent, we
choose as simple a parametrization of the cold gas mass as possible.
Finally, we make the simplest assumption that the mass of all globular
clusters formed in the merger is linearly proportional to the mass of
this cold gas. 

Although such model appears extremely simplistic, we have some
confidence that it may capture main elements of the formation of
massive clusters.  \citet{kravtsov_gnedin05} used a cosmological
hydrodynamic simulation of the Galactic environment at high redshifts
$z > 3$ and found dense, massive gas clouds within the protogalactic
disks.  If the high-density regions of these clouds formed star
clusters, the resulting distributions of cluster mass, size, and
metallicity are consistent with those of the Galactic metal-poor
clusters.  In that model the total mass of clusters formed in each
disk was roughly proportional to the available gas mass, $M_{GC}
\propto M_g$, just as we assume here.

We tune the parameters of our semi-analytical model to reproduce the
metallicity distribution of the Galactic globular clusters, as
compiled by \citet{harris96}.  This distribution is dominated by the
metal-poor clusters but is also significantly bimodal.  We attempt to
construct a model without explicitly differentiating the two modes and
test if bimodality could arise naturally in the hierarchical framework.

We adopt a working definition of red clusters as having $\feh > -1$
and blue clusters as having $\feh < -1$.  This definition should also
roughly apply to extragalactic globular cluster systems.  We use the
concordance cosmology with $\Omega_0 = 0.3$, $\Omega_\Lambda = 0.7$,
$h=0.7$.

\section{Prescription for Globular Cluster Formation}
  \label{sec:outline}

\subsection{Cold Gas Fraction}
\label{sec:coldgasfraction}
We follow the merging process of protogalactic dark matter halos using
cosmological $N$-body simulations of three Milky Way-sized systems
described in \citet{kravtsov_etal04}.  The simulations were run with
the Adaptive Refinement Tree code \citep{kravtsov_etal97} in a $25 \,
h^{-1}$ Mpc box.  Specifically, we use merger trees for three large
host halos and their corresponding subhalo populations.  The three
host halos contain $\sim 10^6$ dark matter particles and have virial
masses $\sim 10^{12}\, \Msun$ at $z=0$.  Two halos are neighbors,
located at 600~kpc from each other.  The configuration of this pair
resembles that of the Local Group.  The third halo is isolated and is
located 2~Mpc away from the pair.  All three systems experience no
major mergers at $z < 1$ and thus could host a disk galaxy like the
Milky Way.

In addition to the host halos, the simulation volume contains a large
number of dwarf halos that begin as isolated systems and then at some
point accrete onto the host halo.  Some of these satellites survive as
self-gravitating systems until the present, while the rest are
completely disrupted by the tidal forces.  We allow both the host and
the satellite systems to form globular clusters in our model.

We adopt a simple hypothesis, motivated by the hydrodynamic simulation
of \citet{kravtsov_gnedin05}, that the mass in globular clusters,
$M_{GC}$, that forms in a given protogalactic system is directly
proportional to the mass of cold gas in the system, $M_g$.  We define
the corresponding mass fraction, $f_g$, of cold gas that will be
available for star cluster formation in a halo of mass $M_h$ as
\begin{equation}
  f_g \equiv {M_g \over f_b \, M_h},
\end{equation}
where $f_b \approx 0.17$ is the universal baryon fraction
\citep{komatsu_etal10}.

The gas fraction cannot exceed the total fraction of baryons accreted
onto the halo, which is limited by external photoheating and depends
on the cutoff mass $M_c(z)$:
\begin{equation}
  f_{in} = {1 \over (1 + M_c(z)/M_h)^{3}}.
  \label{eq:fin}
\end{equation}
We use an updated version of the cutoff mass as a function of redshift
(originally defined by \citealt{gnedin00}), based on our fitting of
the results of recent simulations by \citet{hoeft_etal06},
\citet{crain_etal07}, \citet{tassis_etal08}, and
\citet{okamoto_etal08}:
\begin{equation} 
  M_c(z) \approx 3.6\times 10^9 \, e^{-0.6\, (1+z)} \ h^{-1}\, \Msun.
  \label{eq:charmass}
\end {equation}
Given the scatter in simulation results and the numerical limitations
of the modeling of gas physics, a reasonable uncertainty in this mass
estimate is of the order 50\%.  However, the resulting cluster mass
and metallicity distributions are not very sensitive to the exact form
of this equation.  Note that \citet{orban_etal08} provided an earlier
revision of the equation for $M_c(z)$; our current form is more
accurate.  If $M_c(z)$ falls below the mass of a halo with the virial
temperature of $10^4$ K, we set $M_c(z)$ equal to that mass:
\begin{equation} 
  M_{c, \rm min}(z) = M_4 \equiv
  1.5 \times 10^{10} \Delta_{\rm vir}^{-1/2}\, {H_0 \over H(z)}\
h^{-1}\, \Msun,
\end {equation}
where $\Delta_{\rm vir} = 180$ is the virial overdensity and $H(z)$ is
the Hubble parameter at redshift $z$.  This criterion ensures that we
only select halos able to cool efficiently via atomic hydrogen
recombination lines.

Some of the baryons accreted onto a halo may be in a warm or hot phase
(at $T > 10^4$ K) unavailable for star formation, thus $f_g < f_{in} <
1$.
We assume that only the gas in cold phase ($T \ll 10^4$ K) is likely
to be responsible for star cluster formation.  The cold gas fraction
$f_g$ is calculated by combining several observed scaling relations.
From the results of \citet{mcgaugh05}, the average gas-to-stellar mass
ratio in nearby spiral and dwarf galaxies can be fitted as
\begin{equation}
  \frac{M_g}{M_*} \approx \left({\frac{M_*}{M_s(z)}}\right)^{-0.7},
  \label{eqn:mgms}
\end{equation}
where $M_s$ is a characteristic scale mass, which we found to be
$M_s(z=0) \approx 4 \times 10^9 \, \Msun$.  This relation saturates at
low stellar masses, where $f_g$ cannot exceed $f_{in}$.  At higher
redshift the only information on the gas content of galaxies comes
from the study by \citet{erb_etal06} of Lyman break galaxies at $z =
2$.  These authors estimated the cold gas mass by inverting the
Kennicutt-Schmidt law and using the observed star formation rates.
These estimates are fairly uncertain and model-dependent.  Within the
uncertainties, their results can be fitted by the same formula but
with a different scale mass: \mbox{$M_s(z=2) \approx 2 \times 10^{10}
\, \Msun$}.  To extend this relation to all epochs, we employ a
relation that interpolates the two values:
\begin{equation}
  M_s(z) \approx 10^{9.6 + 0.35 \, z}\ \Msun.
  \label{eqn:scalemass}
\end{equation}

An additional scaling relation is needed to complement equation
(\ref{eqn:mgms}) with a prescription for stellar mass as a function of
halo mass.  We compile it by combining the observed stellar
mass--circular velocity correlation with the theoretical circular
velocity--halo mass correlation.  \citet{woo_etal08} found that the
stellar mass of the dwarf galaxies in the Local Group correlates with
their circular velocities, which are taken as the rotation velocity
for the irregular galaxies or the appropriately scaled velocity
dispersion for the spheroidal galaxies.  In the range $10^7\, \Msun <
M_* < 10^{10}\, \Msun$, appropriate for the systems that may harbor
globular clusters, the correlation is $V_c \propto M_*^{0.27 \pm
0.01}$.  This can be inverted as $M_* \approx 1.6\times 10^9\ \Msun\
(V_c / 100\ {\rm km~s}^{-1})^{3.7}$.

Cosmological $N$-body simulations show that dark matter halos, both
isolated halos and satellites of larger halos, exhibit a robust
correlation between their mass and maximum circular velocity
\citep[e.g., Fig. 6 of][]{kravtsov_etal04}: $V_{\rm max} \approx 100\
(M_h / 1.2 \times 10^{11}\, \Msun)^{0.3}\ {\rm km~s}^{-1}$.  This
maximum circular velocity of dark matter is typically lower than the
rotation velocity of galaxies because of the contribution of stars and
gas. To connect the two velocities, we apply the correction $V_c =
\sqrt{2}
\, V_{\rm max}$, which reflects the observation that the mass in dark
matter is approximately equal to the mass in stars over the portions
of galaxies that contain the majority of stars. Then the equations in
the last two paragraphs lead to $M_* \approx 5.5 \times 10^{10} \,
(M_h/10^{12} \, \Msun)^{1.1}\ \Msun$.

We also need to extend this local relation to other redshifts.
\citet{conroy_wechsler09} matched the observed number densities of
galaxies of given stellar mass with the predicted number densities of
halos of given mass from $z=0$ to $z\sim 2$, averaged over the whole
observable universe.  They find that the stellar fraction $f_*$ peaks
at masses $M_h \sim 10^{12} \, \Msun$ and declines both at lower and
higher halo masses.  The range of masses of interest to us is below
the peak, where we can approximate $f_*$ dependence on halo mass as a
power-law.  The results from Fig. 2 of \citet{conroy_wechsler09} are
best fit by a steeper relation than we derived for the Local Group and
also show significant variation with redshift at lower halo masses
$\sim 10^{11}\, \Msun$: $M_* \propto M_h^{1.5} (1+z)^{-2}$ (there is
much less variation with time around the peak at $10^{12}\, \Msun$,
implying only a weak evolution in the total stellar density at $z<1$).
We adopt the same redshift dependence to our local relation, while
using the shallower slope derived from \citet{woo_etal08} because it
deals with a population of halos in the mass range corresponding to
the Milky Way progenitors.  The stellar mass fraction of isolated
halos is thus
\begin{equation}
  f_* = {M_* \over f_b \, M_h} \approx
     0.32 \left({ \frac{M_h}{10^{12}\, \Msun} }\right)^{0.1} (1+z)^{-2}.
  \label{eq:fstar}
\end{equation}
This relation steepens at low masses because of two additional limits
on the gas and stellar fractions, which we impose to constrain the
range of equations (\ref{eqn:mgms}--\ref{eq:fstar}) to be physical.

First, the sum of the gas and stars (``cold baryons'') cannot exceed
the total amount of accreted baryons in a halo:
\begin{equation}
  f_* + f_g \le f_{in}.
  \label{eq:coldbar}
\end{equation}
At each redshift, there is a transition mass $M_{h,\rm cold}$, below
which $f_* + f_g = f_{in}$ and above which $f_* + f_g < f_{in}$.  For
masses $M < M_{h,\rm cold}$ (but not too low, see next paragraph), we
set $f_{g,\rm revised} = f_{in} - f_*$, with $f_*$ still given by
equation (\ref{eq:fstar}).  We consider the baryons that are not
included in $f_g$ or $f_*$ to be in the warm-hot diffuse phase of the
interstellar medium.

Second, the ratio of stars to cold baryons, $\mu_* \equiv f_*/(f_* +
f_g)$, is not allowed to increase with decreasing halo mass.  For
massive halos ($M_h > M_{h,\rm cold}$) $\mu_*$ monotonically decreases
with decreasing mass because of the condition (\ref{eqn:mgms}).  At
some intermediate masses $M_{h,\mu} < M_h < M_{h,\rm cold}$, $\mu_*$
continues to decrease but the gas fraction is reduced by the condition
(\ref{eq:coldbar}).  At $M_h < M_{h,\mu}$, $\mu_*$ would reverse this
trend and increase with decreasing halo mass because the cold gas is
almost completely depleted.  Such a reversal is unlikely to happen in
real galaxies, which would not be able to convert most of their cold
gas into stars.  Therefore, for all masses $M_h < M_{h,\mu}$ we fix
$\mu_*$ to be equal to the minimum value reached at $M_{h,\mu}$.  This
affects both $f_*$ and $f_g$.

We expect our stellar mass prescription to apply in the range of halo
masses from $10^9$ to $10^{12}\, \Msun$, at least for the Local Group.
However, this relation breaks when a halo becomes a satellite of a
larger system.  Satellite halos often have dark matter in the outer
parts stripped by tidal forces of the host, while the stars remain
intact in the inner parts.  Unless the satellite is completely
disrupted, we keep its stellar mass fixed at the value it had at the
time of accretion, even though the halo mass may subsequently
decrease.

\begin{figure}[t]
\vspace{-0.4cm}
\centerline{\epsfxsize3.5truein \epsffile{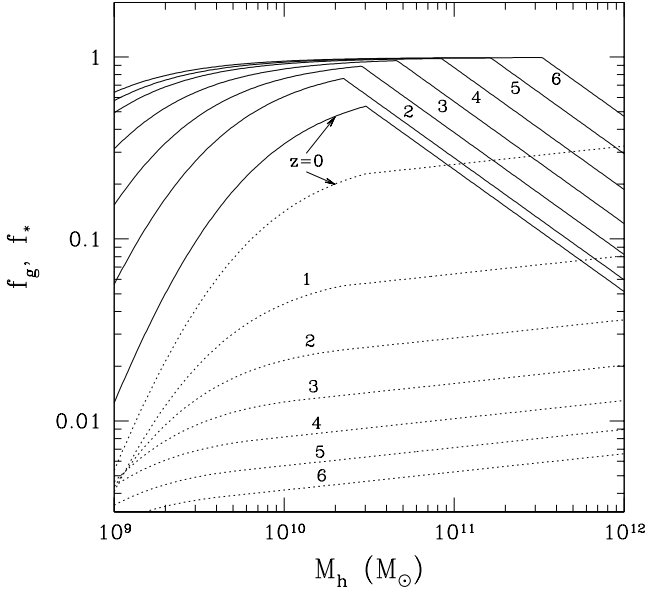}}
\vspace{-0.2cm}
\caption{Gas mass fraction ({\it solid lines}) and stellar mass fraction
  ({\it dotted lines}) vs. halo mass in our model at redshifts
  $z=0,1,2,3,4,5,6$.  Stellar fractions monotonically increase with
  time while gas fractions decrease with time.  The kink in the curves
  is due to our restriction on the maximum stellar fraction via
  eq.~(\ref{eq:coldbar}).}
\vspace{0.3cm}
  \label{fig:mgas}
\end{figure}

The simultaneous effects of the above scaling relations are difficult
to understand as equations.  Figure~\ref{fig:mgas} illustrates
graphically the values of the gas and stellar fractions used in our
model at various cosmic times.  At $z=0$ the gas fraction peaks for
halos with $M_h \sim 3\times 10^{10}\ \Msun$.  At lower masses it is
reduced by the amount of accreted baryons (eq.~\ref{eq:coldbar}),
while at higher masses it is reduced by the gas-to-stars ratio
(eq.~\ref{eqn:mgms}).  The stellar fraction follows equation
(\ref{eq:fstar}) at high masses but drops faster at low masses because
of the constraint (\ref{eq:coldbar}).  For a Galaxy-mass halo, $M_h
\approx 10^{12}\, \Msun$, our model gives $M_* \approx 5.5 \times
10^{10}\, \Msun$ and $M_g \approx 9 \times 10^9\, \Msun$.  These
numbers are consistent with the observed amount of the disk and bulge
stars and the atomic and molecular gas in the Galaxy, from
\citet{binney_tremaine08}.

At earlier epochs at all masses of interest, the gas fraction is
higher and the stellar fraction is lower.  There is a range of halos
with $M_h \gtrsim 10^{10}\ \Msun$, which have an almost 100\% gas
fraction at redshifts $z>3$.  Such halos should be most efficient at
forming massive star clusters.

\begin{figure}[t]
\vspace{-0.4cm}
\centerline{\epsfxsize3.5truein \epsffile{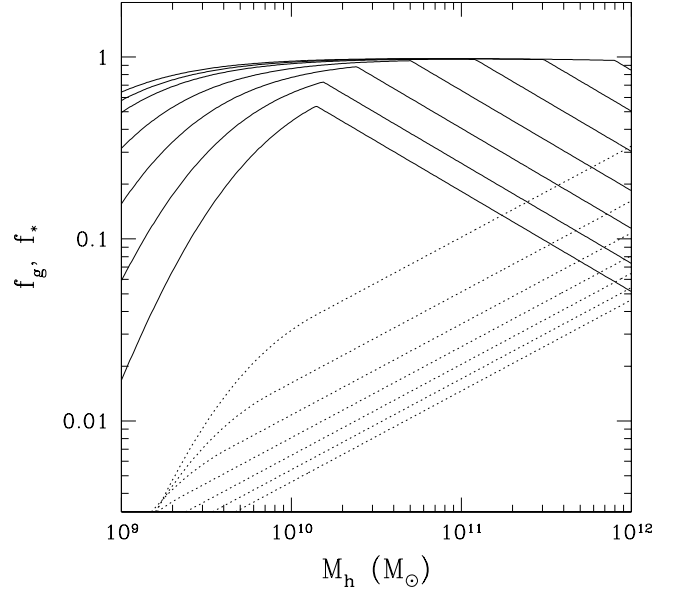}}
\vspace{-0.2cm}
\caption{Same as Fig.~\protect\ref{fig:mgas}, but for an alternative
  stellar mass prescription, eq.~(\ref{eq:fstar1}).}
\vspace{0.3cm}
  \label{fig:mgas1}
\end{figure}

We realize that our adopted relations for the evolution of the stellar
and gas mass are not unique, as we are basing each fit on two data
points.  In order to test the sensitivity of our results to these
assumptions, we consider alternative functional forms for these fits
in Section~\ref{sec:altm}.  In particular, we give the stellar
fraction a steeper dependence on halo mass and weaker dependence on
cosmic time:
\begin{equation}
  f_{*,\rm alt} = 0.32 \left({\frac{M_h}{10^{12}\, \Msun}}\right)^{0.5}
(1+z)^{-1}.
  \label{eq:fstar1}
\end{equation}
Such a slower evolution of the stellar mass is consistent with the
observational studies of \citet{borch_etal06}, \citet{bell_etal07},
and \citet{dahlen_etal07}.  The corresponding gas and stellar
fractions are shown in Figure~\ref{fig:mgas1}.  Note that the amount
of cold gas available for cluster formation is not strongly affected
by this change (compare Figs.~\ref{fig:mgas} and \ref{fig:mgas1}).

\subsection{Rate of Cluster Formation}
  \label{sec:formation_rate}

Having fixed the parametrization of the available cold gas, we then
relate the gas mass of a protogalaxy to the combined mass of all
globular clusters it can form within $\sim 10^8$~yr (the timescale of
the simulation output).  We based it on the rate derived in
\citet{kravtsov_gnedin05}:
\begin{equation}
  M_{GC} = 3 \times 10^6 \, \Msun \left({1 + p_2}\right)
     {M_g/f_b \over 10^{11} \, \Msun}.
  \label{eqn:massGC}
\end{equation}
An additional factor, $1+p_2$, allows us to boost the rate of cluster
formation.  Such a boost may be needed because we form new clusters
only at arbitrarily chosen epochs corresponding to the simulation
outputs.  Unresolved mergers between the outputs may require $p_2 >
0$.  In our model we find the best fit to the Galactic metallicity
distribution for $p_2 \sim 3$ (see Table~\ref{tab:par}).

Note also that equation (\ref{eqn:massGC}) imposes the minimum mass of
a halo capable of forming a globular cluster.  Based on dynamical
disruption arguments (Section~\ref{sec:dyn}) we track only clusters
more massive than $M_{\rm min} = 10^5\, \Msun$.  Since we always have
$M_g < f_b \, M_h$, in order to form even a single cluster with the
minimum mass, the halo needs to be more massive than $10^9\, \Msun$.
For gas-rich systems at high redshift, $M_{GC} \sim 10^{-4}\, M_g/f_b
\sim 10^{-4}\, M_h$.

Given the combined mass of all clusters to be formed in an event,
$M_{GC}$, our procedure for assigning masses to individual clusters is
as follows.  We first draw the most massive cluster, which we call the
nuclear star cluster, even though we do not have or use the
information about its actual location within the host galaxy and it is
not important for our current study.  The mass assigned to the nuclear
cluster, $M_{\rm max}$, is derived from the assumed initial cluster
mass function, $dN/dM = M_0 M^{-2}$:
\begin {equation}
  1 = \int_{M_{\rm max}}^{\infty} \frac{dN}{dM} dM,
\end {equation}
which gives $M_{\rm max} = M_0$.  This normalization is constrained by
the integral cluster mass:
\begin {equation}
  M_{GC} = \int_{M_{\rm min}}^{M_{ \rm max}} M \frac{dN}{dM} dM
         = M_{\rm max} \ln{M_{\rm max} \over M_{\rm min}}.
\end {equation}
The power-law initial mass function agrees both with the observations
of young star clusters and the hydrodynamic simulations.  After the
nuclear cluster is drawn, the masses of smaller clusters are selected
by drawing a random number $0 < r < 1$ and inverting the cumulative
distribution: $r = N(<M)/N(<M_{\rm max})$:
\begin{equation}
  M = \frac{M_{\rm min}}{1 - r \left({1 - M_{\rm min}/M_{\rm
max}}\right)}.
  \label{eq:indmass}
\end{equation}
We continue generating clusters until the sum of their masses reaches
$M_{GC}$.

The formation of clusters is triggered by a gas-rich major merger of
galaxies, which includes mergers of satellite halos onto the main halo
as well as satellite-satellite mergers.  New clusters form when the halo
mass at $i$-th simulation
output exceeds the mass at the previous output by a certain factor,
and at the same time the cold gas fraction exceeds a threshold value:
\begin{equation} 
  {\tt case-1:} \quad M_{h,i} > (1+p_3)\ M_{h,i-1}
                \quad {\rm and}
                \quad f_g > p_4.
\end{equation}
Also, we require that the maximum circular velocity does not decrease
in this time step, to ensure that the mass increase was real rather
than a problem with halo identification.  We have experimented with a
more relaxed criterion for the main halo than for satellite halos,
with $p_{3, \rm main} < p_{3, \rm sat}$, but did not find a
significantly better fit to the mass or metallicity distributions.  We
therefore keep a single value of $p_3$ for all halos.

For some model realizations, we consider an optional alternative
channel for cluster formation without a detected merger, if the cold
gas fraction is very high:
\begin{equation} 
  {\tt case-2:} \quad f_g > p_5,
\end{equation}
where the threshold $p_5$ is expected to be close to 100\%.  This
channel allows continuous cluster formation at high redshift when the
galaxies are extremely gas-rich.  High-redshift galaxies are probably
in a continuous state of major and minor merging, but because of their
lower masses it is more difficult to detect such mergers in the
simulation.  Additional motivation for this channel follows from some
nearby starburst galaxies that are forming young massive clusters
despite appearing isolated.  {\tt Case-2} formation is allowed only for
isolated halos before they are accreted into larger systems and become
satellites.  The epoch of accretion is defined by the last timestep
before the orbit of the subhalo falls permanently within the virial
radius of its host.

\begin{table}
\begin{center}
\caption{\sc Fiducial Model Parameters}
\label{tab:par}
\begin{tabular}{lll}
\tableline\tableline\\
\multicolumn{1}{l}{Parameter} &
\multicolumn{1}{l}{Value} &
\multicolumn{1}{l}{Effect}
\\[2mm] \tableline\\
$\sigma_{\rm met}$ &  0.1  & log-normal dispersion of mass-metallicity
relation\\
$p_2$     &  3.0   & boost of the rate of cluster formation\\
$p_3$     &  0.2   & minimum merger ratio\\
$p_4$     &  0.04  & minimum cold gas fraction for {\tt case-1}
formation\\
$p_5$     &  0.98  & minimum cold gas fraction for {\tt case-2}
formation
\\[2mm] \tableline
\end{tabular}
\end{center}
\vspace{0.1cm}
\end{table}

Our model sample combines clusters formed in the main halo and in its
satellites, either surviving or disrupted.  We exclude clusters from
the satellites that have a galactocentric distance at $z=0$ greater
than 150 kpc, which is the largest distance of a Galactic globular
cluster.  We apply the criteria for cluster formation at every
timestep of the simulation (every $\sim 10^8$ yr) for each of the
three main halos and their satellite populations.  The rate of cluster
formation per every merger event is therefore approximately
$M_{GC}/10^8$ yr.

In order to compare the distribution of clusters obtained from our
analysis to the distribution of Galactic globular clusters, we
normalize the total number of model clusters by the ratio of the
Galaxy mass to the simulated halo masses at $z=0$:
\begin{equation}
  N_{\rm normalized} = N_{\rm model}\ \frac{M_{MW}}{M_{h1} + M_{h2} +
M_{h3}}.
  \label{eq:norm}
\end{equation}
We take $M_{MW} = 10^{12}\, \Msun$ and use $M_{h1} = 2.37 \times
10^{12}\, \Msun$, $M_{h2} = 1.77 \times 10^{12}\, \Msun$, and $M_{h3}
= 1.70 \times 10^{12}\, \Msun$ from \citet{kravtsov_etal04}.

\subsection{Metallicity}
\label{sec:metallicity}
The iron abundance is assigned to each model cluster according to the
estimated average metallicity of its host galaxy.  The latter we
obtain from the mass-metallicity relation for dwarf galaxies of the
Local Group at $z=0$ as formulated by \citet{woo_etal08}:
\begin{equation} 
  \feh_0 = -1.8 + 0.4\log{\left({\frac{M_*}{10^6 \, \Msun}}\right)}.
  \label{eqn:massmetal1}
\end{equation}
In fact, the same fit is valid for the smallest, ultrafaint dwarfs
studied by \citet{kirby_etal08}.  Thus we apply this relation to all
protogalactic systems in our simulation volume.

We also include the evolution of this relation with cosmic time, based
on the available observations of Lyman-break galaxies at $z \approx 2$
\citep{erb_etal06}, Gemini Deep Deep Survey galaxies at $z \approx 1$
\citep{savaglio_etal05}, and cosmological hydrodynamic simulations
that provide the average metallicity of galaxies \citep{brooks_etal07,
dave_etal07}:

\begin{equation} 
  \feh(t) \approx \feh_0 - 0.03 \left({t_0 - t \over 10^9 \ {\rm
yr}}\right).
  \label{eqn:massmetal2}
\end{equation}
While this temporal evolution is probably real, it can change the
metallicity for the same stellar mass by at most 0.36 dex in 12 Gyr.
This amount is smaller than the 0.4 dex change of $\feh$ due to the
stellar mass variation by a factor of 10.  In our model, globular
clusters form in protogalaxies with a range of stellar masses of
several orders of magnitude (see Fig.~\ref{fig:histopanel} below).

The observed mass-metallicity relation for a large sample of galaxies
observed by the SDSS has an intrinsic scatter of at least 0.1 dex
\citep[e.g.,][]{tremonti_etal04}.  We account for it, as well as for
possible observational errors, by adding a Gaussian scatter to our
calculated $\feh$ abundances with a standard deviation of $\sigma_{\rm
met} = 0.1$ dex.  The exact value of this dispersion is not important
and can go up to 0.2 dex without affecting the results significantly.

Using equations (\ref{eqn:massmetal1}) and (\ref{eqn:massmetal2})
along with the procedures of Section~\ref{sec:formation_rate}, we can
generate a population of star clusters with the corresponding masses
and metallicities.  The model contains two random factors: the scatter
of metallicity and the individual cluster masses assigned via equation
(\ref{eq:indmass}).  We sample these random factors by creating 11
realizations of the model with different random seeds.  Each
realization combines clusters in all three main halos.  Taking into
account that the halos are about twice as massive as the Milky Way,
the expected number of clusters in each model realization is $\sim 150
\, (\mathrm{the\ observed\ number}) \times (2.37+1.77+1.7) \approx
870$.  The total set of all 11 realizations includes $\sim 9500$
clusters.  For the purpose of conducting statistical tests on the
distributions of cluster mass and metallicity, we consider each
realization separately and then take the median value of the
calculated statistic.

For convenience, we provide a list of the most important equations we
used in the model in table  \ref{tab:equations}.
\begin{table}
\begin{center}
\caption{\sc Summary of Model Equations}
\label{tab:equations}
\begin{tabular}{lll}
\tableline\tableline\\
\multicolumn{1}{l}{Equation} &
\multicolumn{1}{l}{Section} &
\multicolumn{1}{l}{Description}
\\[2mm] \tableline\\
\ref{eq:fin} & \ref{sec:coldgasfraction} & fraction of baryons accreted onto a halo\\
\ref{eq:charmass} & \ref{sec:coldgasfraction} & cutoff mass for baryon accretion\\
\ref{eqn:mgms} & \ref{sec:coldgasfraction} & cold gas mass relative to stellar mass\\
\ref{eq:fstar} & \ref{sec:coldgasfraction} & stellar mass relative to halo mass\\
\ref{eqn:massGC} & \ref{sec:formation_rate} & mass in globular clusters relative to gas mass\\
\ref{eqn:massmetal1} & \ref{sec:metallicity} & stellar mass-metallicity relation for halos\\
\ref{eq:mt} & \ref{sec:dyn} & evolution of cluster mass
\\[2mm] \tableline
\end{tabular}
\end{center}
\vspace{0.1cm}
\end{table}

\section{Dynamical Disruption}
  \label{sec:dyn}

Star clusters are prone to gradual loss of stars, and in some cases,
total disruption by internal and external processes.  It is expected
that the mass function of globular clusters has evolved through cosmic
time, from an initial (probably, a power law) distribution to the
approximately log-normal distribution that is observed today.  Since
the main focus of this paper is on the observable properties of the
Galactic population, we evolve all of our model clusters dynamically
from their time of formation until the present epoch.  We adopt the
evaporation via two-body relaxation and stellar evolution as the
mechanisms for mass loss.  Tidal shocks are ignored for simplicity.
Cluster mass changes because of the decrease of the number of stars,
$N_*(t)$, by evaporation and the decrease of the average stellar mass,
$\mav(t)$, by stellar evolution:
\begin{equation}
  {1\over M}{dM \over dt} \equiv 
    {1\over N_*}{dN_* \over dt} + {1\over \mav}{d\mav \over dt}
    = -\nu_{\rm ev}(M) -\nu_{\rm se}(t) {\mav(0) \over \mav}.
  \label{eq:dmdt}
\end{equation}
We have assumed, as done in the recent literature, that the
evaporation rate depends only on cluster mass. The time $t$ for each
cluster is measured from the moment of its formation.

We adopt the calculation of \citet{prieto_gnedin08} for the
time-dependent mass-loss rate due to stellar evolution, $\nu_{\rm
se}(t)$ (see their Fig. 7).  That calculation uses the relation
between star's initial mass and remnant mass from
\citet{chernoff_weinberg90} and the main-sequence lifetimes from
\citet{hurley_etal00}.  Over time, stellar evolution reduces the
cluster mass by up to 40\%, for a \citet{kroupa01} IMF.  This implies
that no clusters are disrupted by stellar evolution alone, and the net
effect is only a shift in the mass distribution towards the lower end.

We now need to derive the evaporation rate, $\nu_{\rm ev}(M)$, as a
function only of cluster mass.  We begin by writing down the standard
approximation \citep{spitzer87} using the half-mass relaxation time,
$t_{rh}$:
\begin{equation}
  \nu_{\rm ev} = \frac{\xi_{e}}{t_{rh}} 
           = \frac{7.25 \, \xi_{e} \, \mav \, G^{1/2} \, \ln\Lambda}
                  {M^{1/2} \, R_{\rm h}^{3/2}},
 \label{eqn:massev}
\end{equation}
where $\xi_{e}$ is the fraction of stars that escape per relaxation
time, $R_{\rm h}$ is the half-mass radius, and $\ln\Lambda$ is the
Coulomb logarithm.  We take $\overline{m} = 0.87 \, \Msun$ for a
Kroupa IMF, and $\ln\Lambda = 12$, which is a common value used for
globular clusters \citep{spitzer87}.

We then assume that at the time of formation $R_{\rm h}$ depends only
on cluster mass, as $R_{\rm h} \propto M^{\delta_0}$, and not on the
position in the host galaxy.  As a fiducial model, we use a constant
density model where $\delta_0 = 1/3$ \citep{kravtsov_gnedin05,
prieto_gnedin08}.  The relation for the initial size is normalized
with respect to the median observed mass of Galactic clusters, $2
\times 10^5\, \Msun$, and their median size of 2.4 pc:
\begin{equation}
  R_{\rm h}(0) = 2.4 \ \mathrm{pc} 
                 \left({M(0) \over 2\times 10^5\
\Msun}\right)^{\delta_0}.
  \label{eqn:Rhinitial}
\end{equation}
A similar relation extends to other dynamically hot stellar systems:
nuclear star clusters and ultracompact dwarf galaxies
\citep{kissler-patig_etal06}.  The mass-size relation may change over
the course of the cluster evolution.  We consider a power-law relation
with a potentially different slope, so that the half-mass radius
responds to changes in the cluster mass as
\begin{equation}
 \label{eqn:Rht}
  {R_{\rm h}(t) \over R_{\rm h}(0)} = \left({M(t) \over
M(0)}\right)^{\delta}.
\end{equation}
Our preferred value is again $\delta = 1/3$, but we also discuss results
for other choices of $\delta_0$ and $\delta$.  Recent $N$-body models
of cluster disruption are consistent with $\delta \approx 1/3$
\citep{trenti_etal07, hurley_etal08}.  Note that cluster sizes are
only used as an intermediate step in the derivation of $\nu_{\rm
ev}(M)$ and can be subsequently ignored.  The evaporation time thus
becomes
\begin{equation}
  \nu_{\rm ev}^{-1} \approx 10^{10}\, {\rm yr}\,
     \left({\xi_{e} \over 0.033}\right)^{-1}
     \left({M(0) \over 2\times 10^5\,
\Msun}\right)^{\frac{1+3\delta_0}{2}}
     \left({M(t) \over M(0)}\right)^{\frac{1+3\delta}{2}}.
  \label{eq:nuev}
\end{equation}

\begin{figure}[t]
\centerline{\epsfxsize3.3truein \epsffile{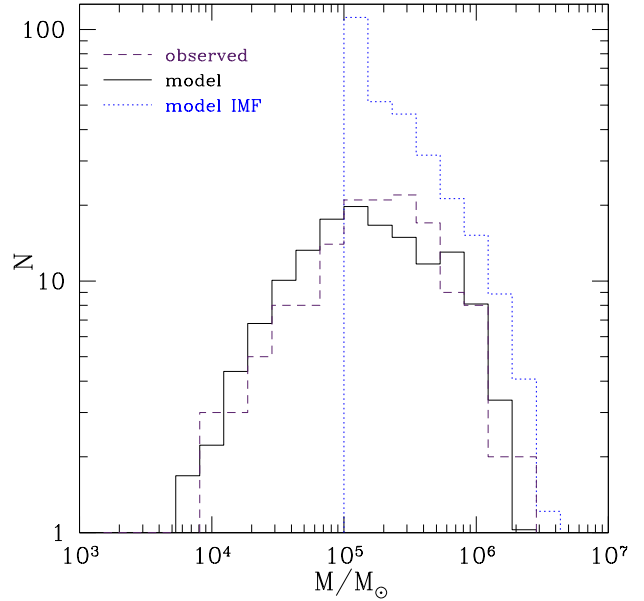}}
\caption{Dynamically evolved clusters at $z=0$ in the fiducial model
  with $\xi_e = 0.033$, $\delta=\delta_0=1/3$ ({\it solid histogram}),
  compared to the observed distribution of Galactic globular clusters
  ({\it dashed histogram}).  Dotted histogram shows the combined
  initial masses of model clusters formed at all epochs, including
  those that did not survive until the present.  In the model we do
  not follow clusters with the initial masses below $10^5\, \Msun$.}
  \label{fig:mf}
\end{figure}

The fraction $\xi_{e}$ is not well constrained.  The lower limit on
$\xi_{e}$ is achieved in isolated clusters, for which $\xi_{e} =
0.0074$ \citep{ambartsumian38, spitzer40}.  Tidally-truncated clusters
lose stars at a faster rate, as first calculated by \citet{henon61}
and \citet{spitzer_chevalier73}.  Using orbit-averaged Fokker-Planck
models of cluster evolution, \citet{gnedin_etal99} found $\xi_{e}$
varying between 0.02 and 0.08 depending on time and cluster
concentration (their Fig. 4 and Table 2).  More recently, realistic
direct $N$-body models became possible \citep[e.g.,][]{baumgardt01,
baumgardt_makino03}.  These calculations revealed that the gradual
escape of stars through the tidal boundary, which is not spherical as
in the Fokker-Planck calculations, breaks the linear scaling of the
disruption time with the relaxation time.  \citet{baumgardt01}
suggested that the evaporation time scales as $\nu_{\rm ev}^{-1}
\propto t_{rh}^{3/4}$.  \citet{gieles_baumgardt08} verified this
relation and found almost no dependence on the cluster half-mass
radius.  Instead, they proposed an explicit dependence on the
Galactocentric distance $R_G$ and velocity $V_G$, to reflect the
strength of the local tidal field: $\nu_{\rm ev}^{-1} \propto
\omega^{-1} \equiv R_G/V_G$.  This gives $\nu_{\rm ev}^{-1} \propto
M^{3/4} \omega^{-1}$.  Their formula is similar to the empirical
estimates of the disruption time by \citet{lamers_etal05}: $\nu_{\rm
ev}^{-1} \propto M^{0.65}$.

Since the calculation of the local tidal field is currently beyond our
simple model, we ignore the dependence on the Galactocentric distance
but argue that we can incorporate the result of
\citet{gieles_baumgardt08} for the disruption timescale by using a
lower value of $\delta_0 = \delta = 1/9$.  With this choice of the
exponents, our equation (\ref{eq:nuev}) gives $\nu_{\rm ev}^{-1}
\propto M^{2/3}$.  We discuss these alternative models in Section
\ref{sec:altdyn}.

For consistency with \citet{prieto_gnedin08}, we adopt $\xi_{e} =
0.033$ for the fiducial model.

With the above ingredients, we can now compute the cluster mass at
time $t$ after formation by inverting $\nu_{\rm ev}(M)$ in equation
(\ref{eq:dmdt}) and assuming that most of the stellar evolution mass
loss
happens much faster than the evaporation:
\begin{equation}
  M(t) = M(0) \; \left[{ 1-\int_0^t \nu_{\rm se}(t') \, dt' }\right] \; 
       \left[{ 1 - {1+3\delta \over 2} \nu_{\rm ev,0}\, t
}\right]^{2/(1+3\delta)},
  \label{eq:mt}
\end{equation}
where $\nu_{\rm ev,0} \equiv \nu_{\rm ev}(M=M(0))$.

\begin{figure}[t]
\centerline{\epsfxsize3.3truein \epsffile{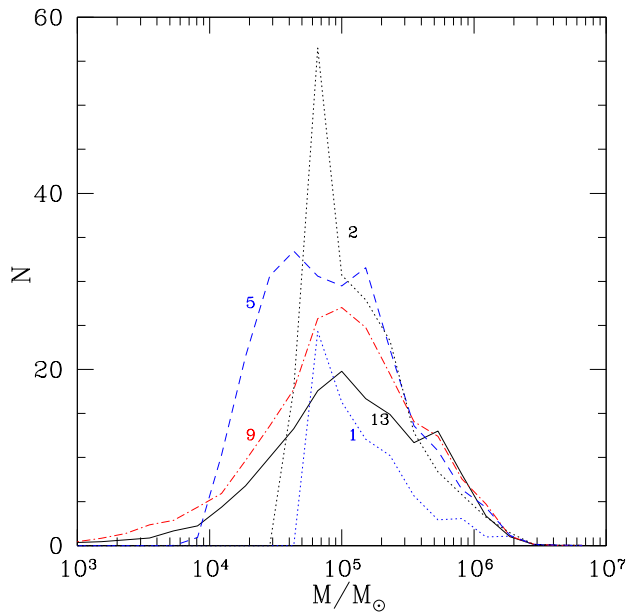}}
\caption{The mass function of clusters in the fiducial model at
  different epochs corresponding to the cosmic times of 1 Gyr ($z
  \approx 5.7$, {\it dotted}), 2 Gyr ($z \approx 3.2$, {\it dotted}),
  5 Gyr ($z \approx 1.3$, {\it dashed}), 9 Gyr ($z \approx 0.5$, {\it
  dot-dashed}), and 13.5~Gyr ($z=0$, {\it solid}).}
  \label{fig:gc_massevo}
\end{figure}

The initial mass function of globular clusters is evolved from the
time of formation until the present epoch and is shown in
Figure~\ref{fig:mf} for the fiducial model.  The observed mass
function in the Milky Way is well represented by a log-normal
distribution.  We derive the masses of the Galactic clusters by taking
their absolute V-band magnitudes from the \citet{harris96} catalog and
assuming a constant mass-to-light ratio $M/L_V = 3\,
\Msun/\mathrm{L}_{\odot}$.  The functional form of a Gaussian built
around $\log{M}$ for the observed sample is given by
\begin{equation}
  {dN \over d\log{M}} = {1 \over \sqrt{2\pi}\sigma_M}
      \exp{\left[-{(\log{M}-\logMav)^2 \over 2\sigma_M^2}\right]},
\end{equation} 
with the mean $\logMav = 5.22$ and standard deviation $\sigma_M =
0.61$, in solar masses.  The predicted mass function in the fiducial
model with $\xi_e=0.033$ and $\delta=\delta_0=1/3$ is consistent with
the observations.  The Kolmogorov-Smirnov (KS) test probability of the
two mass functions being drawn from the same distribution is $P_{KS,M}
= 7.4\%$.  This value is the median of the KS probabilities for the 11
random realizations of the model.  The model distribution is also well
fit by a Gaussian, with $\logMav = 5.14$ and $\sigma_M = 0.65$.
 
The mean of the model distribution is slightly lower than observed,
implying that the disruption process needs to be stronger to fully
reconcile with the data.  Clusters that start out with low mass but
are not disrupted effectively over their lifetime over-populate the
low end of the present-day model mass function.  Old and
intermediate-age clusters that started with initial mass $5 < \log{M}
< 5.4$ and survived until the present era appear to be the main cause
of this discrepancy.

Figure~\ref{fig:gc_massevo} illustrates the evolution of the mass
function over cosmic time as an interplay between the continuous
buildup of massive clusters ($M > 10^5\, \Msun$) and the dynamical
erosion of low-mass clusters ($M < 10^5\, \Msun$).  Since we do not
track the formation of clusters below $M_{\rm min}$, the low end of
the mass function was built by a gradual evaporation of more massive
clusters.  The strongest bout of cluster formation happens between a
cosmic time of 1 and 5 Gyr, and a peak of the mass function forms at
$M \sim 3 \times 10^4 \, \Msun$.  The peak moves to larger masses,
$\sim 10^5 \, \Msun$ by $t = 9$ Gyr, while a power-law tail develops
at low masses.  A significant fraction of low-mass clusters is
disrupted between 9 Gyr and the present, as few new clusters are
produced.

\begin{figure}
\centerline{\epsfxsize3.3truein \epsffile{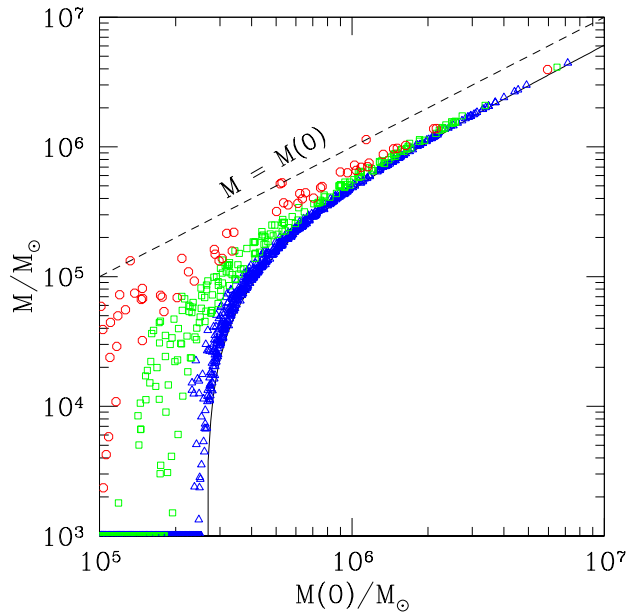}}
\caption{Final mass of model clusters vs. their initial mass, for a
  single realization of the fiducial model.  Clusters are divided into
  three age groups: {\it triangles} represent old clusters (age > 10
  Gyr), {\it squares} represent intermediate age clusters (5 Gyr < age
  < 10 Gyr), and {\it circles} represent young clusters (age < 5 Gyr).
  All disrupted clusters are placed at the bottom of the plot, to
  illustrate the range of their initial mass.  The birthline of
  clusters, $M=M(0)$, is plotted as a dashed line for reference.}
  \label{fig:massconnect}
\end{figure}

The relation between the cluster initial and final masses is shown in
Figure~\ref{fig:massconnect}.  Old clusters that have undergone
significant amounts of dynamical and stellar evolution form a tight
sequence on this plot.  The lower boundary with a dense concentration
of points corresponds to the expression $M = 0.63 \, [M(0) - 2.6\times
10^5\ \Msun]$, which reflects 13 Gyr of stellar and dynamical
evolution according to equation (\ref{eq:mt}) with the fiducial values
of the parameters.  Thus an old cluster must have an initial mass of
at least $2.6\times 10^5\, \Msun$ to survive dynamical disruption.
Clusters in the younger age groups fill the space between their
birthline and this boundary.  The youngest clusters have the
shallowest slope at low mass, as few of them have had enough time to
undergo significant disruption.  The mean final mass for all three age
groups is about the same, implying that some of the oldest globular
clusters could have been more massive at the time of their formation
than clusters that have formed recently in the local universe.

\citet{fall_zhang01} suggested that a low-mass end of the mass
function should approach $dN/dM \approx const$ as a result of dominant
disruption by two-body evaporation.  Our mass function in the range
$3.5 < \log(M/\Msun) < 5.0$ is consistent with a power law
$\log(dN/d\log{M}) = 0.89 \log{M} - 3.04$, or $dN/dM \propto
M^{-0.11}$, in good agreement with the expectation.

\begin{figure}[t]
\centerline{\epsfxsize3.3truein \epsffile{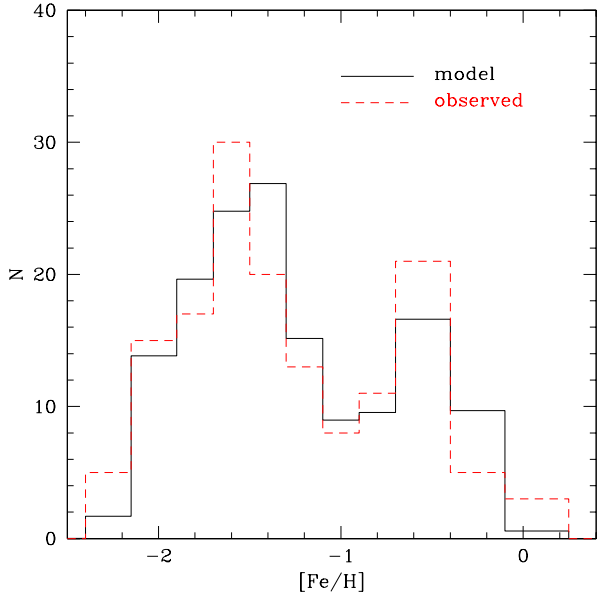}}
\caption{Metallicities of model clusters formed at all epochs that
  have survived dynamical disruption by $z=0$ in the fiducial model
  ({\it solid histogram}), compared to the observed distribution of
  Galactic globular clusters ({\it dashed histogram}).}
  \label{fig:gc_metal}
\vspace{0cm}
\end{figure}

\section{Results}

\subsection{Exploration of the Parameter Space}

Overall, our model has five adjustable parameters
(Table~\ref{tab:par}).  To explore possible degeneracies among these
parameters, and to find the parameter set that produces the
best-fitting metallicity distribution, we set up a grid of models in
which each of the parameters was varied within a finite range of
values.  The range was taken to be large enough to explore all
physically relevant values of each parameter.

The boost for cluster formation, $p_2$, varied from 0 to 5.  For
consistency with the rate derived in the hydrodynamic simulation of
\citet{kravtsov_gnedin05}, we aimed to keep this parameter at low
values.

The minimum mass ratio for mergers, $p_3$, varied between 0.15 and 0.5.
It is consistent with typical major merger criteria used in the
literature (e.g. \citealt{beasley_etal02} use $p_3 = 0.3$).

The cold gas fraction required for cluster formation during a merger,
$p_4$, could be relatively low but non-zero, so that we considered $0
< p_4 < 0.2$.  This threshold parameter accounts for why disk galaxies
like the Milky Way are still forming stars despite a low gas fraction,
while ellipticals are not.

The gas fraction for {\tt case-2}, $p_5$, has to be very high -- above
90\%, as our prescription predicts that many halos have a very high
gas fraction at high redshift and could over-produce blue globular
clusters (as was the case in the \citealt{beasley_etal02} model).

We considered several values for $\sigma_{\rm met}$ but found that a
value of 0.2 or higher smeared out the peaks in the metallicity
distribution, while a value of 0 failed to fill the extreme ends of
the distribution.  We therefore include only three values in our
search, $\sigma_{\rm met} = 0, 0.1, 0.2$.

\begin{figure}[t]
\centerline{\epsfxsize3.3truein \epsffile{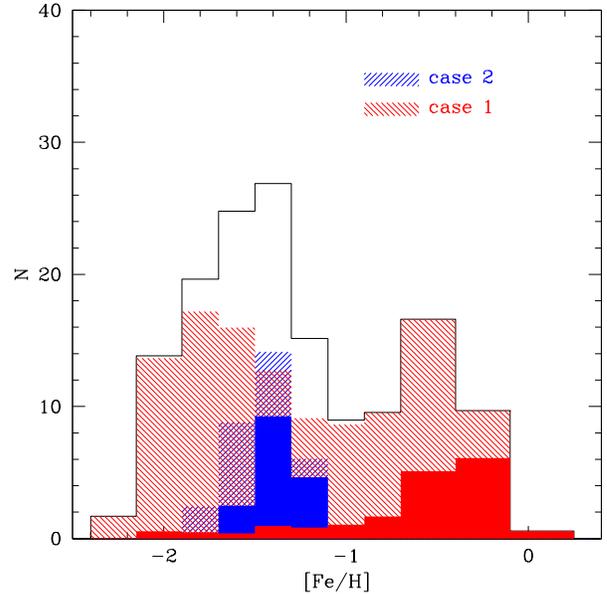}}
\caption{Metallicity distribution in the fiducial model, split by the
  formation criterion: major mergers ({\tt case-1}) and early mergers
  ({\tt case-2}).  Solid histograms show the clusters formed in the main
  Galactic halo.}
\vspace{0cm}
  \label{fig:gc_met_case}
\end{figure}

We find the best-fit model by searching through the multi-parameter
space and maximizing the KS probabilities of the metallicity
distribution, $P_{KS,Z}$, and the mass function, $P_{KS,M}$, being
consistent with observations.  The likelihood function also contains
additional factors that force the parameters towards the values that
we consider ideal.  We require the model to produce the observed
number of clusters, $N \approx 150$, scaled by the host galaxy mass as
in equation (\ref{eq:norm}).  We wish to maximize the fraction of
clusters formed in the main disk, $f_{\rm disk}$, to be consistent
with the observed spatial distribution (Section~\ref{sec:spatial}).
We penalize the likelihood function for large values of $p_{2}$ and
for any young clusters formed after $t=10$ Gyr, $N_{\rm after10}$.  We
also wish to minimize the fraction of clusters formed through the {\tt
case-2} channel, $f_{\rm case2}$, for simplicity of the model.
Finally, we want to increase the likelihood of the metallicity
distribution being bimodal, as characterized by the Dip test, $P_{\rm
dip}$, which we discuss later in Section~\ref{sec:bimodality}.  The
actual likelihood function that we maximize is given by
\begin{eqnarray}
  \log{\cal L} &=& 
     \log{P_{KS,Z}} + 0.3\log{P_{KS,M}} -[(N-150)/30]^2 +\log{f_{\rm
disk}}
      \nonumber\\ 
     && - 0.15 p_2 - 20 N_{\rm after10}/N - 0.4 f_{\rm case2} + 3
\log{P_{\rm dip}}.
  \label{eqn:like}
\end{eqnarray}
The coefficients for each term were adjusted heuristically until we
found that their relative weights matched our expectation to select
acceptable distributions.  The ``best-fit'' distribution that maximizes
${\cal L}$ is therefore a subjective fiducial model that we use to
illustrate how the bimodality may arise.  We then look at how many
model realizations are similar to the ``best-fit'' for other possible
values of the parameters.

\begin{figure}[t]
\centerline{\epsfxsize3.3truein \epsffile{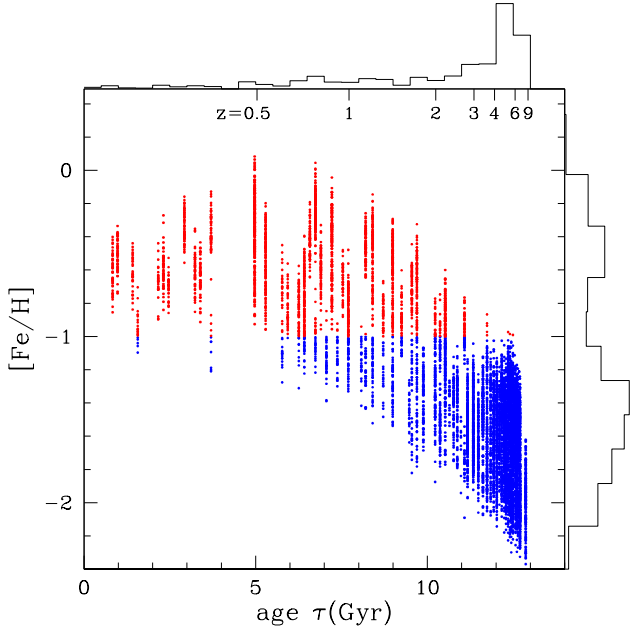}}
\caption{Age-metallicity relation in all 11 realizations of the
  fiducial model ($\approx 9500$ clusters).  The build-up of massive
  halos drives the steep slope of this relation at early epochs.
  Outer histograms show marginalized distributions on linear scale.
  Notice an order-of-magnitude spread in metallicities of clusters
  forming at a given epoch.}
\vspace{0.3cm}
  \label{fig:agemet}
\end{figure}

\subsection{Age and Metallicity Distributions}

Figure~\ref{fig:gc_metal} shows the predicted best-fit metallicity
distribution of model clusters and the observed distribution of
Galactic globular clusters, both metal-poor and metal-rich.  Note that
we require our model to have the same formation criteria for both
cluster populations; we do not explicitly differentiate between the
two modes.  The only variable is the gradually changing amount of cold
gas available for star formation.  Yet, the model predicts two peaks
of the metallicity distribution, centered on $\feh = -1.54$ and $\feh
=-0.58$, in remarkable agreement with the observations.  The standard
deviation of the red peak is 0.24 dex and of the blue peak is 0.32
dex.

The probability of KS test of the model and data samples being drawn
from the same distribution is $P_{KS,Z} = 80\%$, that is, they are
fully consistent with each other.  The number of surviving clusters is
$N=147$, also matching the observations.  Even though our current
model is extremely simple, this bimodality is reproduced naturally,
without explicit assumptions about truncation of the production of
metal-poor clusters at some early epoch or about the formation of
metal-rich clusters in a merger of two spiral galaxies.

We find that the main halo contributes more significantly to the red
peak than it does to the blue peak (Figure \ref{fig:gc_met_case}).  In
particular, clusters with the highest $\feh$ appear to have been
formed primarily by late merging into the main halo.

The fraction of clusters formed via {\tt case-2} channel is $f_{\rm
case2} = 22\%$.  These clusters produce a single-peaked distribution
of blue clusters.  In contrast, clusters formed in major mergers
contribute to both red and blue modes, in about equal proportions.  We
return to this point in the discussion of globular cluster systems of
elliptical galaxies in Section~\ref{sec:discussion}.

Clusters that formed after $z=2$ constitute the bulk of the red peak
and contribute little to the blue peak in the metallicity distribution
(Figure \ref{fig:agemet}).  The strength of this result implies that
the gas reservoir and the rate of hierarchical merging at intermediate
redshifts is conducive to the creation of red clusters.  This result
lends itself well to the idea that the simulation of
\citet{kravtsov_gnedin05} was only able to reproduce the metal-poor
population of globular clusters because the simulation was stopped at
$z \approx 3$.

Our prescription links cluster metallicity to the average galaxy
metallicity in a one-to-one relation, albeit with random scatter.
Since the average galaxy metallicity grows monotonically with time,
clusters forming later have on the average higher metallicity.  The
model thus encodes an age-metallicity relation, in the sense that
metal-rich clusters are younger by several Gyr than their metal-poor
counterparts.  This relation is required in the model to reproduce the
observed metallicity distribution, because very old galaxies cannot
produce high enough metallicities.  However, Figure~\ref{fig:agemet}
shows that clusters of the same age may differ in metallicity by as
much as a factor of 10, as they formed in the progenitors of different
mass.

\begin{figure}[t]
\centerline{\epsfxsize3.4truein \epsffile{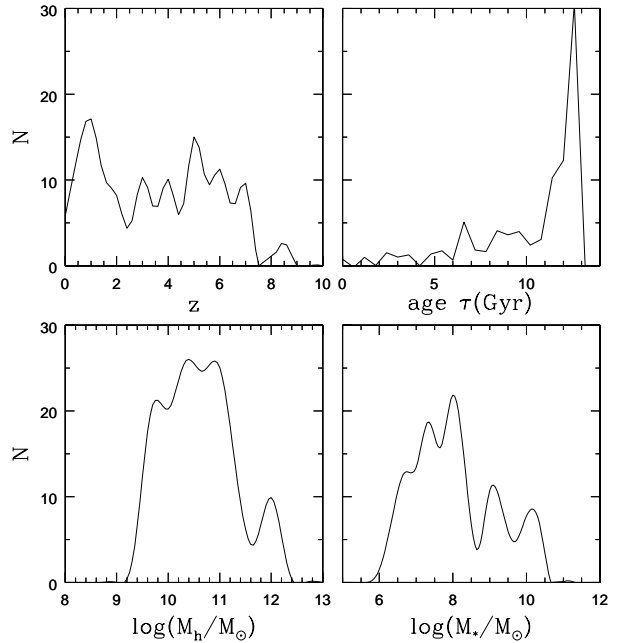}}
\caption{Number of clusters in the fiducial model as a function of the
  environment: redshift of formation ({\it top left panel}), present
  age ({\it top right panel}), host halo mass at the time of formation
  ({\it bottom left panel}), and host stellar mass ({\it bottom right
  panel}).}
\vspace{0.3cm}
  \label{fig:histopanel}
\end{figure}

Available observations of the Galactic globular clusters do not show a
clear age-metallicity relation, but instead indicate an age spread
increasing with metallicity \citep{deangeli_etal05,
marin-franch_etal09, dotter_etal10, forbes_bridges10}.  Red clusters
have younger mean age overall and may be as young as $\tau \approx
7$~Gyr.  Our model does not appear to be in an obvious conflict with
this trend.  We define cluster age as $\tau \equiv t_0 - t_f$, where
$t_f$ is the time of formation.  We find the mean age of 11.7 Gyr for
the blue population and 6.4 Gyr for the red population, with the
standard deviation of 1.3 Gyr and 2.7 Gyr, respectively.  More
accurate dating of the Galactic and extragalactic clusters is needed
to falsify the predicted age-metallicity trend.

Distributions of the cluster formation time and environment in the
fiducial model are shown in Figure~\ref{fig:histopanel}.  The age
distribution, which peaks strongly between 11 and 13 Gyr, demonstrates
that the majority of our clusters is still very old and falls in line
with the observed perception of globular clusters.  However, the
distribution of formation redshift appears remarkably flat in the
range $1<z<7$, emphasizing that the clusters were not formed in a
single event but rather through the continuous process of galaxy
formation.  Few clusters were formed prior to the era of reionization,
as sufficiently large quantities of gas could not be condensed to meet
the mass threshold for cluster formation at redshifts $z>9$.  The
distributions of the total and stellar mass of the host galaxies
extend over three orders of magnitude.  Their extended high-mass tails
contribute to the strength of the red peak, as the most massive halos
would form most metal-rich clusters.

\begin{figure}[t]
\centerline{\epsfxsize3.3truein \epsffile{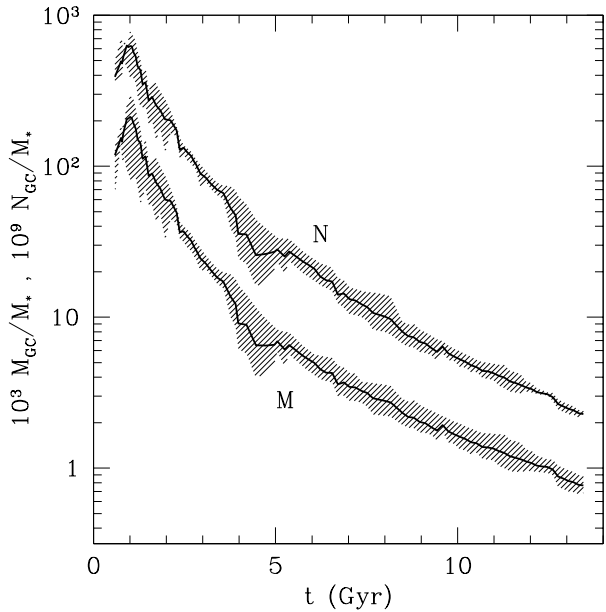}}
\caption{Ratios of total cluster mass and number to the galaxy stellar
  mass, summed over all systems that are located within 150 kpc of
  the center at $z=0$.  Lines represent the average over all three
  host halos and their corresponding subhalo populations.  Shading
  represents the scatter in these ratios, given by the lowest and
  highest values among the three hosts.  Massive star cluster
  formation was a much more dominant component of galactic star
  formation at early times than it has been for the last 10 Gyr.}
\vspace{0.3cm}
  \label{fig:MGCMstar}
\end{figure}

Globular clusters form much earlier than the majority of field stars.
Figure~\ref{fig:MGCMstar} shows the fraction of galaxy stellar mass
locked in massive star clusters, normalized for convenience as $10^3
\, M_{GC}/M_*$.  To calculate this ratio, we summed over all
protogalactic systems that would end up within 150 kpc of the galaxy
center at $z=0$, regardless of their location at earlier times.  Thus
it represents a global cluster formation efficiency in a Milky
Way-sized environment.  Specific realizations of the model differ in
detail in the three host halos, by as much as a factor of 2.  This
scatter is shown by the shaded region on the plot.  The globular
cluster mass includes their continuous formation and the mass loss due
to the dynamical evolution.  A striking prediction of the model is a
very high cluster fraction at early times, near $t=1$ Gyr, of
$M_{GC}/M_* \approx 10-20\%$.  Star cluster production may have been a
dominant component of galactic star formation at $z>3$.  By $t=3$ Gyr
($z \approx 2$), the cluster fraction drops to only a few percent, as
expected for a galaxy undergoing active star formation.  At the
current epoch, massive star clusters make up less than 0.1\% of the
stellar mass.  The predicted ratio is progressively more uncertain at
higher redshift because it relies on our extrapolated prescription for
the galactic stellar mass.  The low-redshift prediction should be
robust.  We also show a variant of the specific frequency parameter
related to the number of clusters, $T \equiv N/(M_*/10^9 \, \Msun)$,
introduced by \citet{zepf_ashman93}.  It shows a similar decline with
time, reaching $T \approx 2$ at the present.

These global cluster formation efficiencies agree with many
observations across galaxy types.  \citet{rhode_etal05} find $T \sim
1$ for both red and blue clusters in the field and group spiral
galaxies.  This parameter increases with the galaxy mass.  In the
Virgo cluster, \citet{peng_etal08} find $T \sim 5$ for galaxies in the
mass range appropriate for the Milky Way.  \citet{mclaughlin99}
estimated the cluster mass fraction in both spiral and elliptical
galaxies to be $M_{GC}/(M_* + M_g) \approx 0.0026 \pm 0.0005$.  This is
larger than what we find by a factor of several, but we count in $M_*$
all stars out to 150 kpc, which includes some satellite galaxies as well
as the host.  Therefore, both predicted cluster efficiencies at $z=0$
are reasonable.  Their rise at high redshift is an interesting
prediction of the model.

The model also shows that the globular cluster system overall is more
metal-poor than the stars in disrupted satellites, which are expected
to form a stellar spheroid of the Galaxy.  We calculated the
mass-weighted metallicity of stars formed in the disrupted satellites
of all three main halos (using eqs. \ref{eq:fstar}, \ref{eq:coldbar},
\ref{eqn:massmetal1}, \ref{eqn:massmetal2}).  This calculation bears
all the uncertainty of our extrapolated time evolution of the stellar
fraction and mass-metallicity relation, but nevertheless provides a
useful estimate.  We find the tail of halo star metallicities as low
as the most metal-poor globular clusters, but the overall stellar
distribution peaks around $\feh \approx -0.3$.  A very similar
situation is observed in NGC 5128 and discussed by \citet{harris10}.
In our model, majority of globular clusters form before the bulk of
field stars and therefore acquire lower metallicities.  For
comparison, the metallicity of stars in surviving satellite galaxies
peaks around $\feh \approx -0.8$ and forms an intermediate population
between the clusters and the field.

Despite our attempts to incorporate it as a major penalty in the
likelihood statistic, we were unable to completely eliminate the
phenomenon of young massive star clusters.  Interestingly, these
clusters did not originate in the main galactic disks.  All clusters
younger than 5 Gyr formed in satellite halos in the mass range $\sim
10^{10} - 10^{11}\, \Msun$, at distances $40-100$~kpc from the
center.  Although the proper sample of the Galactic globular clusters
does not contain any young clusters, there are several young massive
clusters in M31 whose ages were confirmed both from the visual and UV
colors \citep{fusipecci_etal05, rey_etal07} and from the
integrated-light spectroscopy \citep{puzia_etal05}.  The actual
analogs of young model clusters may be found in the LMC, which hosts
globular clusters with a wide range of ages and continues to form
clusters now.  There may even exist young star clusters with masses
$\sim 10^5\, \Msun$ in the Galactic disk, hidden behind tens of visual
magnitudes of extinction but revealing themselves through free-free
emission of their ionization bubbles \citep{murray_rahman10}.  Massive
star cluster formation at late times thus paints a picture consistent
with the idea that today's super star clusters are destined to become
observationally equivalent to globular clusters, as envisioned by
\citet{ashman_zepf92} and \citet{harris_pudritz94}.

\begin{figure}[t]
\centerline{\epsfxsize3.3truein \epsffile{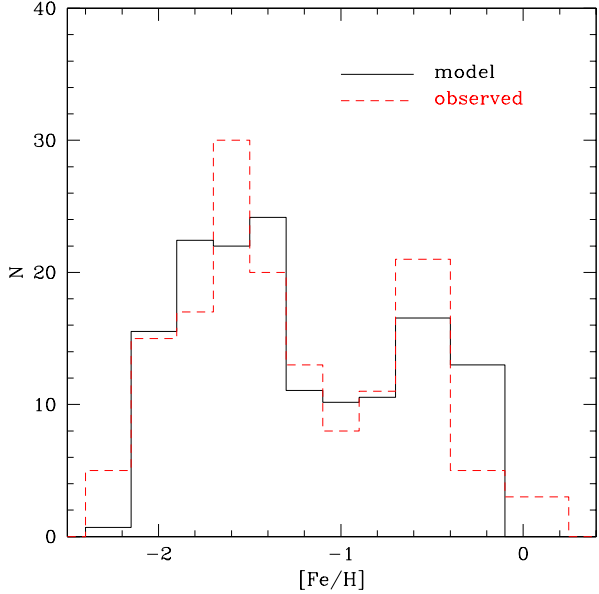}}
\caption{Metallicity distribution at $z=0$ in the model without {\tt
  case-2} formation ({\it solid histogram}), compared to the observed
  distribution of Galactic globular clusters ({\it dashed
  histogram}).}
\vspace{0.3cm}
  \label{fig:gc_metal4}
\end{figure}

A separate criterion for the formation of clusters in extremely
gas-rich systems ({\tt case-2}) is not necessary for achieving a good
fit to the observed metallicity distribution.  Though we feel that the
inclusion of {\tt case-2} formation channel in the model is both
useful and physically motivated, it takes away from the elegance of
using only resolved mergers as a lone formation mechanism.  It turns
out that the main benefit of allowing clusters to form via {\tt
case-2} is seen in the mass function of surviving clusters.  The
high-mass end of the mass function matches the observations better if
massive halos (primarily the main halo) are allowed to form as many
clusters as possible at early times.

We searched the model grid without {\tt case-2}, by setting $p_5=1$,
and found an almost equally good metallicity distribution as in the
fiducial model.  Figure~\ref{fig:gc_metal4} shows that this
distribution also appears bimodal and completely consistent with the
data.  The KS probability is $P_{KS,Z} = 92\%$.  In fact, even the
mass function is only marginally less consistent, $P_{KS,M}=2.0\%$
vs. 7.4\% in the fiducial model.  The parameters used to obtain this
distribution were: $p_2 = 2.85$, $p_3 = 0.16$, $p_4 = 0.04$, $p_5 =
1$, $\sigma_{\rm met} = 0.1$.

\begin{figure}[t]
\centerline{\epsfxsize3.5truein \epsffile{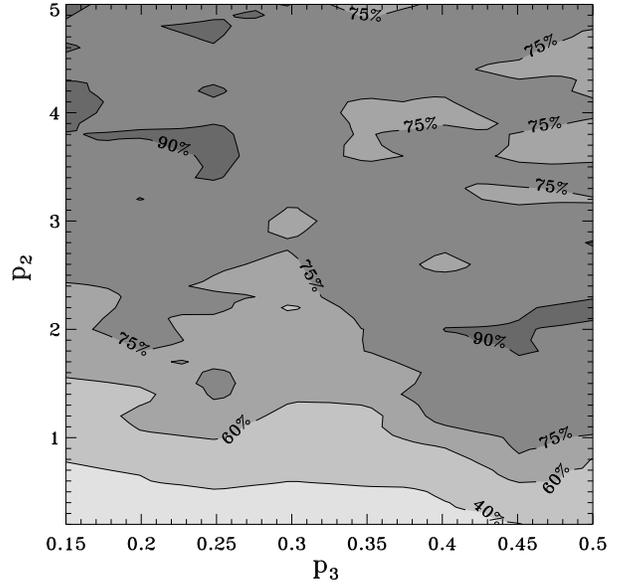}}
\caption{Contour plot of the KS probability for the metallicity
  distributions in the plane of parameters $p_2 - p_3$.  Contour
  labels are the actual probability values, $P_{KS,Z}$.  This plot
  shows that KS test alone cannot rule out any region of the parameter
  space from being statistically consistent with the data.}
\vspace{0.3cm}
  \label{fig:p2p3contourKS}
\end{figure}

\subsection{Sensitivity to Model Parameters}
  \label{sec:degen}

The fiducial distribution discussed above is not unique among our
results in its ability to match the observations.  Significant
degeneracy exists among combinations of the model parameters that
produce metallicity distributions consistent with the Galactic sample.
Many models within the grid have sufficiently high KS probabilities.
In this section we explore which regions of the parameter space
produce models similar to our best fit.

First, let us motivate the use of the likelihood function given by
equation (\ref{eqn:like}) as opposed to using a standard statistical
test to select the best fit.  In the early stages of development of
our model, we relied on KS test alone to help us understand the range
of parameters that produce metallicity distributions that match the
observations.  However, once the model was completed, it became
apparent that KS test alone was not powerful enough for analysis of
the results.  This is clearly demonstrated in Figure
\ref{fig:p2p3contourKS}, which shows the value of the KS probability
$P_{KS,Z}$ as a function of $p_2$ and $p_3$ across their respective
ranges in the grid.  Each point represents the maximum possible
$P_{KS,Z}$ for the given values of $p_2$ and $p_3$ with the other
parameters free to vary within the grid.  This is done to best
represent the full extent of the 5-dimensional parameter space within
a 2-dimensional slice.  Statistically, any distribution with $P_{KS,Z}
> 10\%$ cannot be ruled out with confidence, implying that almost the
entire range of our parameters can produce statistically consistent
distributions!  In addition, although some regions of the parameter
space have higher values of $P_{KS,Z}$ than others, there is no clear
pattern in the contours to help us understand the required physics of
star cluster formation within our semi-analytical recipe.

\begin{figure}[t]
\centerline{\epsfxsize3.5truein \epsffile{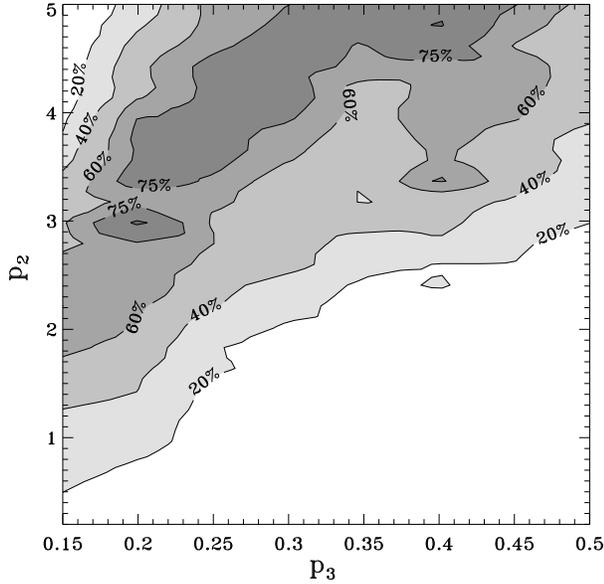}}
\caption{Contour plot of the likelihood statistic ${\cal L}$ in the
  plane of parameters $p_2 - p_3$.  Contour labels show percentages of
  the maximum.  The highest-value region is a degeneracy along the
  line $p_2 = 19 p_3 - 0.91$.}
\vspace{0.3cm}
  \label{fig:p2p3contour}
\end{figure}

\begin{figure}[t]
\centerline{\epsfxsize3.5truein \epsffile{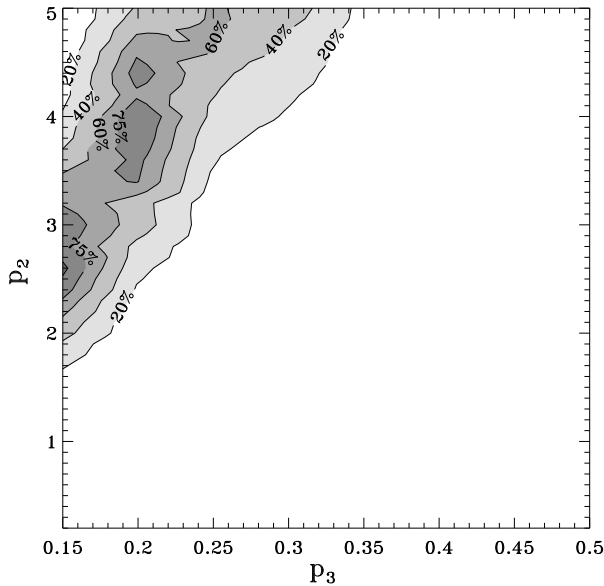}}
\caption{Same as Figure~\ref{fig:p2p3contour} but for the models
  without {\tt case-2}. The highest-value region is a degeneracy along
  the line $p_2 = 24 p_3 - 1.1$.}
  \label{fig:p2p3contour_nocase2}
\end{figure}

In comparison, Figure~\ref{fig:p2p3contour} shows contours of the
value of the likelihood function from equation (\ref{eqn:like}), using
the same scheme described above to maximize the value at each point.
The shape of these contours demonstrates that $p_2$ and $p_3$ are
degenerate in their ability to produce good distributions.  The
degeneracy can most easily be understood by noting that these
parameters directly affect the total number of clusters: $p_2$
controls the cluster formation rate per merger, while $p_3$ selects
eligible mergers.  It is therefore expected that the contours show a
correlation at high levels of the likelihood function, as the
statistic depends sensitively on the total number of clusters.

Figure \ref{fig:p2p3contour_nocase2} shows the same type of contours
as Fig.~\ref{fig:p2p3contour}, but only for distributions with $p_5 =
1$.  Disallowing the {\tt case-2} channel reduces the range of the
parameter space where good distributions are found.  In particular,
compared to the previous plot, Fig.~\ref{fig:p2p3contour_nocase2}
lacks any viable models with $p_3 > 0.3$.  Given the tight and steep
correlation in this plot, it is likely that larger values of $p_3$
would require very high $p_2 > 5$, which may violate current
observational constraints on the cluster formation efficiency.
However, Fig.~\ref{fig:gc_metal4} demonstrates that a good model can
still be found with reasonably small values of $p_2$ and $p_3$,
without the {\tt case-2} channel.

\begin{figure}[t]
\centerline{\epsfxsize3.5truein \epsffile{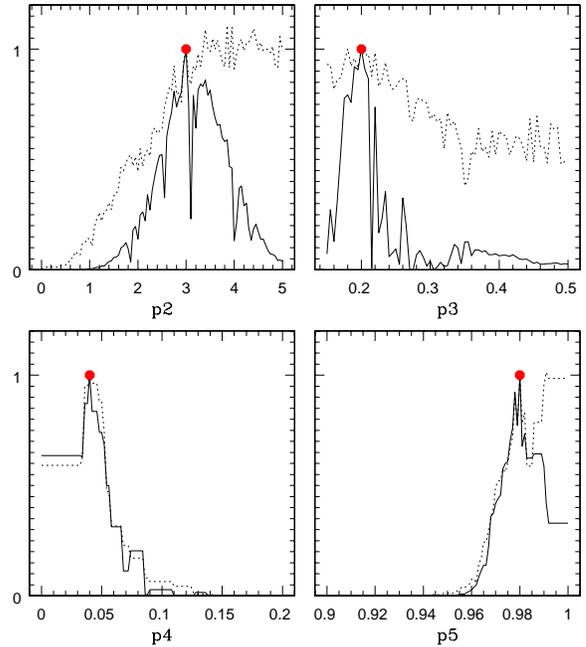}}
\caption{Marginalized single-parameter likelihood distributions around
  the fiducial model, ${\cal L}/{\cal L}_{\rm max}$ ({\it solid
  lines}).  Dashed lines show the metallicity probability $P_{KS,Z}$
  normalized to the fiducial model value.  Compared with $P_{KS,Z}$
  alone, the likelihood function ${\cal L}$ significantly tightens the
  constraints on the best values of the parameters.  Filled circles
  show the fiducial model.}
\vspace{0.3cm}
  \label{fig:deg1d}
\end{figure}

To understand the sensitivity of the likelihood function to individual
parameters, we also considered one-dimensional slices of the parameter
space around the fiducial model, this time allowing for only one
parameter to vary at a time.  Figure~\ref{fig:deg1d} illustrates how
the sharp peaks of ${\cal L}$ allow us to select the best model more
accurately than on the basis of $P_{KS,Z}$ alone.  Particularly as a
function of $p_2$ and $p_3$, $P_{KS,Z}$ varies slowly over the entire
range of the grid.  On the other hand, $p_4$ and $p_5$ must stay
within a small range of their fiducial values in order to achieve
acceptable values of either $P_{KS,Z}$ or ${\cal L}$.

\begin{figure}[t]
\vspace{-0.3cm}
\centerline{\epsfxsize3.4truein \epsffile{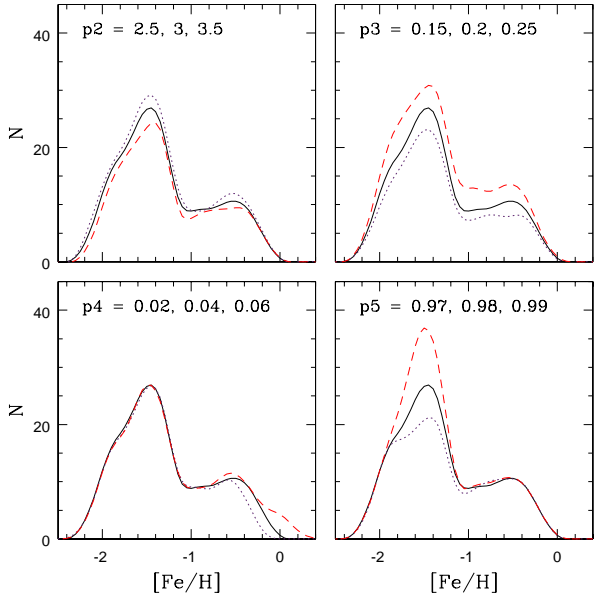}}
\caption{The effects of varying individual model parameters on the
  metallicity distribution.  In each panel a single parameter is
  increased ({\it dotted line}) and decreased ({\it dashed line})
  relative to the fiducial model ({\it solid line}).  The parameter
  values are indicated inside the panels.}
\vspace{0.3cm}
  \label{fig:paramvary}
\end{figure}

Figure~\ref{fig:paramvary} shows variation of the metallicity
distribution when individual parameters deviate from their fiducial
values.  Each parameter can change the shape of the metallicity
function and the number of clusters.  The effects of varying $p_2$ and
$p_3$ are almost opposite, reflecting the degeneracy in the likelihood
contours.  In particular, smaller $p_3$ accommodates more minor
mergers, which allow massive hosts form more metal-rich clusters as
well as some metal-poor clusters.  Decreasing $p_5$ allows more
clusters form through the {\tt case-2} channel; most of such clusters
are metal-poor.  The major role of $p_4$ appears to govern the extent
of the most metal-rich clusters -- lower threshold gas fraction allows
clusters to form in the later, more enriched environments of massive
hosts.

\begin{figure}[t]
\centerline{\epsfxsize3.3truein \epsffile{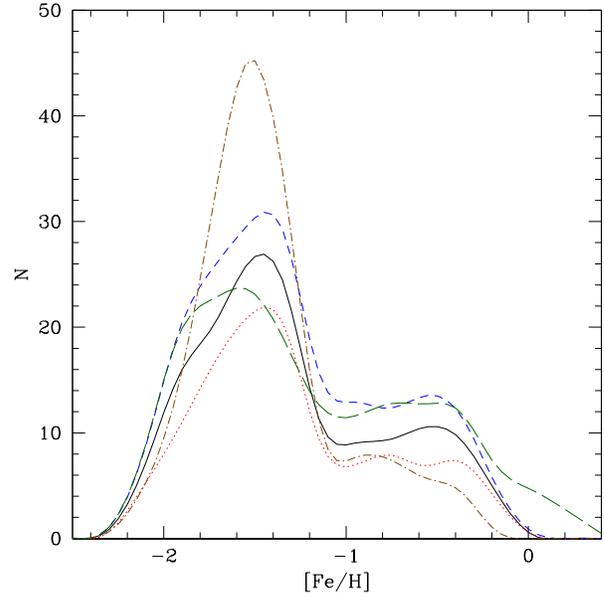}}
\caption{The effects of simultaneous variation of several model
  parameters.  The fiducial model ({\it solid line}) is plotted
  alongside four distributions that illustrate other outcomes of
  our model.  {\it Short-dashed} and {\it dotted lines}
  correspond to respectively lowering (to 0.15) and raising $p_3$ (to
  0.3) away from its fiducial value (0.2).  {\it Long-dashed line}
  corresponds to the model with $p_3=0.15$, $p_4=0$, $p_5=1$.  {\it
  Dot-dashed line} corresponds to the model with $p_3=0.3$,
  $p_4=0.08$, $p_5=0.96$.}
\vspace{0.3cm}
  \label{fig:paramvaryhist}
\end{figure}

Figure~\ref{fig:paramvaryhist} illustrates the response of the
metallicity distribution to simultaneous variations of model
parameters.  First, we plot two distributions where we changed $p_3$
to 0.15 and 0.3 while keeping the other parameters fixed.  The width
of the metal-poor peak broadens as $p_3$ is lowered, indicating that a
wider range of halos in the early universe were able to produce
clusters.  Raising $p_3$ has the opposite effect.  Note that the
locations of the two peaks are remarkably robust to these changes.
Staying at $p_3 = 0.15$, we set $p_5=1$ to eliminate the {\tt case-2}
channel and set $p_4=0$ to allow even gas-poor massive halos at low
redshift to form clusters.  The result (long-dashed line) is a
distribution with a much broader metal-rich peak, which extends well
past the maximum metallicity of the fiducial model.  The dot-dashed
line represents a corresponding change to $p_4 = 0.08$ and $p_5=0.96$
for the $p_3=0.3$ model.  In this case, the metal-rich peak is
severely depleted and remains only as an extended tail of a
single-peaked, metal-poor distribution dominated by {\tt case-2}
clusters.  These distributions are just some of the realizations of
our model that were rejected due to their low values of ${\cal L}$.
All of them have features that conflict with the observed metallicity
distribution in the Galaxy.

\subsection{Origin of the Metallicity Bimodality}
  \label{sec:bimodality}

The KS statistic measures the overall consistency of the model and
observed metallicity distributions, but not specifically bimodality or
multimodality within the distributions.  In order to address the
particular issue of modality, we employ two additional statistical
tests, described in Appendix.

The Gaussian Mixture Modeling test indicates that the fiducial
distribution is bimodal at a high level of significance (better than
0.1\%).  The peak metallicities of both modes and their widths are
close to the observed values and agree with them within the errors.
Both samples easily appear bimodal to the eye because the modes are
well separated, with the dimensionless peak separation ratio $D>3$
(see eq.~\ref{eq:dpeak}).  However, as we discuss in Appendix, the GMM
test is sensitive to the assumption of Gaussian modes.  It may
indicate highly statistically significant split into two modes when
the distribution is truly unimodal but skewed.  For faster and more
robust model selection we consider another test of multimodality.

The Dip test compares the cumulative input distribution with the
best-fitting unimodal distribution.  The maximum distance between the
two corresponds to a dip in the differential distribution.  The Dip
test for the observed Galactic clusters indicates that the
distribution is 90\% likely to not be unimodal.  When applied to our
fiducial model, the Dip test implies it is 99\% likely to not be
unimodal.  However, there is a caveat that the probability of the Dip
test depends on the number of objects in the sample, similarly to KS
test.  The higher significance of the model result does not mean that
the model is actually more bimodal than the data, because we used all
11 random realizations of the model as a combined sample to evaluate
the Dip test.  While this is not a fair comparison to the data, it
allows us to differentiate efficiently among alternative models.

\begin{figure}[t]
\centerline{\epsfxsize3.5truein \epsffile{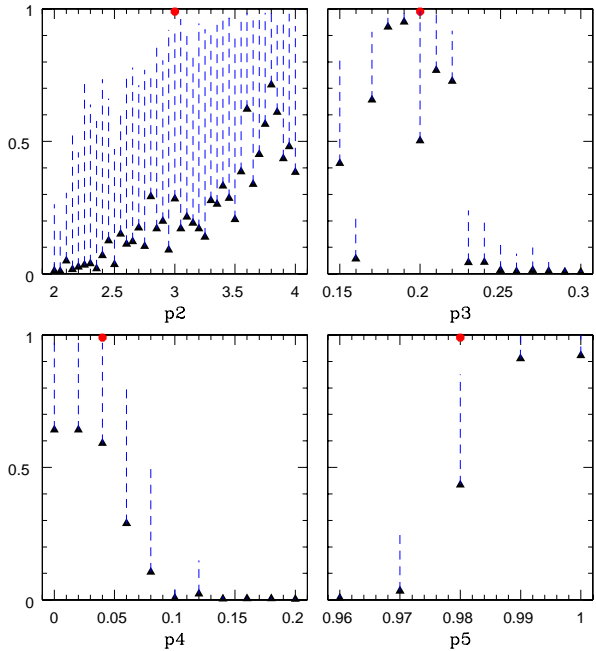}}
\caption{Median values of the Dip probability ({\it triangles}) among
  distributions with $140<N<160$, binned according to each parameter.
  Dashed lines extend to the 75\% quartiles of $P_{\rm dip}$.  The
  fiducial model is shown by a red dot.}
\vspace{0.3cm}
  \label{fig:dip1d}
\end{figure}

We ran the Dip test for all models on the grid in a manner similar to
the likelihood statistic.  The most interesting result of the Dip
statistic comes from one-dimensional slices of the parameter space.
Considering only models with the normalized number of clusters in the
range $140<N<160$, we binned the distributions according to the values
of the four parameters and found the median and quartiles of $P_{\rm
dip}$ in each bin.  Figure~\ref{fig:dip1d} shows several trends. (i)
Distributions with low formation rate $p_2$ are unlikely to be
bimodal.  The 75th percentile of $P_{\rm dip}$ increases
systematically with $p_2$ in the range $2<p_2<3$, but plateaus for
$p_2>3$.  (ii) The most bimodal distributions require $p_3$ to be
small enough to allow for merger ratios 1:5 to trigger cluster
formation.  Between $p_3=0.2$ and $p_3=0.5$, the lower $p_3$ the
better.  However, mass ratios lower than 1:6 may dilute bimodality.
(iii) The gas fraction threshold $p_4$ should be under 10\% for ideal
bimodality, to include mergers of massive galaxies.  (iv) The fraction
$p_5$ has to be close to 1, implying that {\tt case-2} negatively
affects bimodality.  A conclusion from this plot is that bimodality
appears in a significant number of model realizations, for a wide
range of parameters.  At the same time, a similarly large number of
realizations are unimodal.

The metallicity distribution is bimodal if metal-rich clusters
constitute a significant subset of all clusters.  Thus, the fraction
of red clusters, $f_{\rm red}$, is a simple proxy for bimodality.
Indeed, we find a strong correlation between $f_{\rm red}$ and $P_{\rm
dip}$.  The red fraction follows similar, but weaker, trends with
model parameters to those shown in Fig.~\ref{fig:dip1d}.  The median
red fraction correlates most strongly with $p_5$, increasing from
$f_{\rm red} = 16\%$ for $p_5=0.96$ to $f_{\rm red} = 32\%$ for
$p_5=1$.  The red peak is significantly stronger without {\tt case-2}
clusters.

We note that the Dip test, unlike KS test, does not depend on
comparing the model distribution to the Galactic sample.  Therefore,
the trends for bimodality derived from the Dip test should apply to
other globular cluster systems.  We anticipate that bimodality would
likely arise if we applied our model with the parametric constraints
stated above to any cosmological $N$-body simulation that follows the
mass assembly history of a large galaxy.  Further discussion of
applying our model in different galactic environments follows in
Section~\ref{sec:discussion}.

In order to investigate the underlying cause of bimodality, we
examined various properties of merger events.  A merger event is
defined as any time in a halo's track when it meets the criteria for
{\tt case-1} formation.  An important requirement here is the minimum
mass of cold gas needed to produce a cluster that would survive
dynamical disruption.  Through equation (\ref{eqn:massGC}), a cluster
mass $M > 2 \times 10^5\, \Msun$ requires $M_g > 3 \times 10^8\,
\Msun$.  This constraint significantly reduces the number of eligible
mergers.  We considered the distributions of halo mass, lookback time,
and metallicity (without additional dispersion) for all relevant
merger events.  We find that relatively few mergers happen in the
space of high metallicity, high mass, and late time.  Almost half of
the mergers (44\%) take place before $\tau = 12$ Gyr, and only 24\% of
the mergers happen in the last 10 Gyr.  If we also counted the events
that led to now-disrupted clusters, these numbers would spread even
further to 53\% and 17\%, respectively.  Nevertheless, the recent
mergers stand out for two reasons: each such event creates more
clusters, and these younger clusters have better chance of surviving
the dynamical disruption than the older clusters.  Since the number of
clusters formed in each merger is positively correlated with the
galaxy mass, the few stochastic super-massive mergers with high
metallicity are likely to produce a significant number of clusters,
which would separate the red peak from the blue peak.

\begin{figure}[t]
\centerline{\epsfxsize3.3truein \epsffile{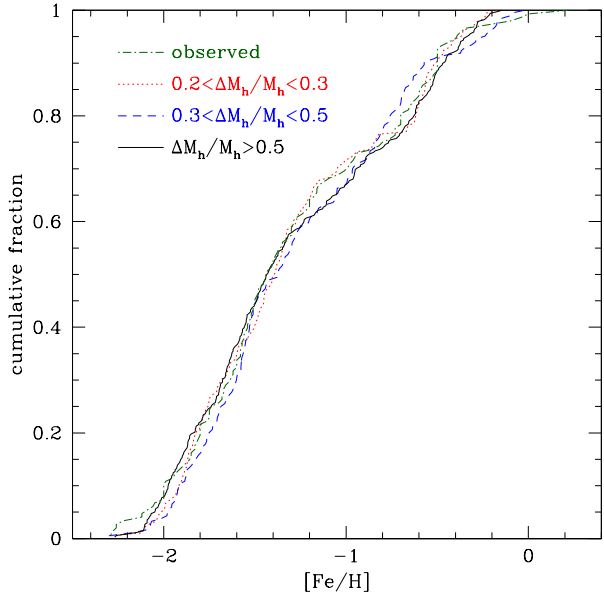}}
\caption{Cumulative metallicity distributions of the fiducial model
  clusters split by the range of merger mass ratios of their formation
  event: $0.2 < \Delta M_h/M_h < 0.3$ ({\it dots}), $0.3 < \Delta
  M_h/M_h < 0.5$ ({\it dashes}), and $\Delta M_h/M_h > 0.5$ ({\it
  solid line}).  The similarity of all three distributions implies
  that metallicity bimodality is not caused by mergers with any
  particular range of mass ratios.  The observed distribution is
  plotted in {\it dot-dashes} for comparison.}
\vspace{0.3cm}
  \label{fig:cumuMetalMerger}
\end{figure}

We also considered that cluster bimodality may be linked to the mass
ratios in the merger events.  "Major" and "minor" mergers have been
proposed to play different roles in galaxy formation, so it is
conceivable that different types of star clusters may be formed
depending on the merger ratio.  In Figure~\ref{fig:cumuMetalMerger} we
plot cumulative metallicity distributions for clusters grouped
according to the mass ratios in the cluster-forming merger event.
Mergers with the closest masses, $\Delta M_h/M_h > 0.5$, contribute
48\% of {\tt case-1} clusters, while the lower two mass ranges each
contribute equal portions of the rest.  Running KS test on these
distributions revealed that they formally represent statistically
different populations.  However, there is no clear-cut range of
metallicities where one type of merging is exclusively producing all
of the clusters, and the overall shapes of the distributions are
similar.  This uniformity suggests that bimodality is a natural
consequence of hierarchical cluster formation regardless of the exact
definition of a "major" merger.

\begin{figure}[t]
\centerline{\epsfxsize3.5truein \epsffile{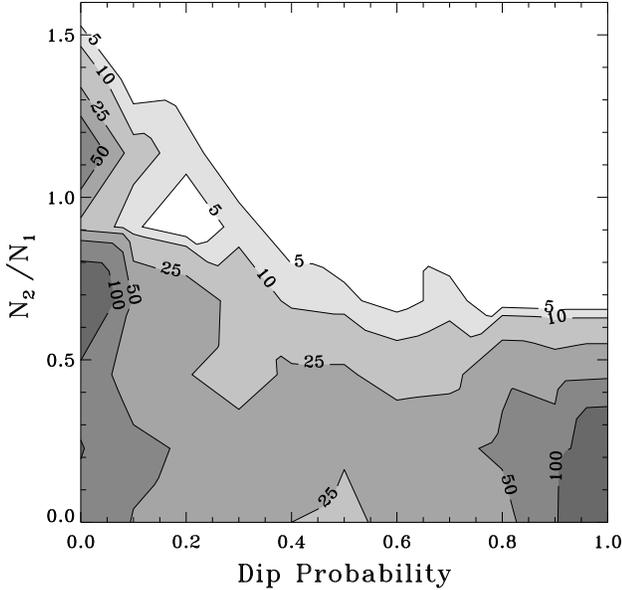}}
\caption{Number of models resulting in particular values of the Dip
  probability and the ratio of {\tt case-2} to {\tt case-1} clusters,
  for all realizations of the parameter grid with the normalized
  number of clusters $140<N<160$.}
\vspace{0.3cm}
  \label{fig:dipcase}
\end{figure}

Figure~\ref{fig:dipcase} shows how many models from the grid fall into
particular ranges of the Dip probability and the ratio of {\tt case-2}
clusters to {\tt case-1} clusters, $N_2/N_1$.  The models are
restricted to have the normalized number of clusters $140 < N_1+N_2 <
160$.  The region with the highest density of models is in the
lower-right corner of the plot, corresponding to high $P_{\rm dip}$
and low $N_2/N_1$.  Low values of $P_{\rm dip}$ are not significant
since they cannot reject a unimodal distribution.  Effectively,
bimodality requires $N_2/N_1 \lesssim 0.5$.  At the significance level
of $P_{\rm dip} = 90\%$, corresponding to the observed distribution,
38\% of the grid models are bimodal if $N_2/N_1 < 0.3$.  This fraction
drops to only 15\% for $0.3 < N_2/N_1 < 1$, and then further to 9\%
for $N_2/N_1 > 1$.  These statistics confirms that bimodal populations
appear only when the {\tt case-2} channel is a secondary formation
mechanism.

Another part of the explanation of bimodality of the surviving
clusters is due to the dynamical evolution.  Most of the disrupted
clusters were old and blue.  If we add these disrupted clusters to the
metallicity histogram in the fiducial model, the blue peak rises by a
factor of 2.  The red peak remains virtually unaffected, since the
more recently formed red clusters are less subjected to dynamical
disruption.  We ran the Dip test on all grid distributions, including
both surviving and disrupted clusters.  Among the models with the
number of surviving clusters in the range $100 < N < 200$, few
distributions have $P_{\rm dip} > 50\%$ and none has $P_{\rm dip} >
80\%$.  This means virtually no bimodality.  Indeed, the distributions
appear almost entirely unimodal, with the peak in the blue end and
nothing more than a tail in the red end.  This leads to a prediction
that late-type galaxies, which have more continuous cluster formation
than early-type galaxies, may be less likely to exhibit bimodal
cluster populations.

\subsection{Alternative Formation Prescriptions}
  \label{sec:altm}

As alluded to in Section \ref{sec:outline}, some equations that we
used in the prescriptions for the stellar mass and the cold gas
fraction were based on only a few observed points.  Currently there is
limited observational or theoretical understanding of how these
functions should behave at high redshift, which is the period of
primary interest for our study.  Below we consider some alternatives
for these prescriptions.

\begin{figure}[t]
\centerline{\epsfxsize3.5truein \epsffile{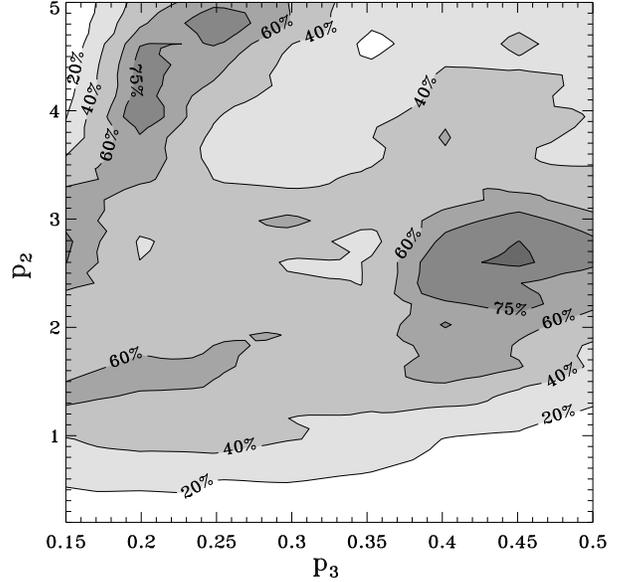}}
\caption{Same as Figure~\ref{fig:p2p3contour} but for the models with
  an alternative prescription for $M_*$ as a function of $M_h$, given
  by equation (\ref{eq:fstar1}).}
\vspace{0.3cm}
  \label{fig:p2p3contour_altmstar}
\end{figure}

\begin{figure}[t]
\centerline{\epsfxsize3.3truein \epsffile{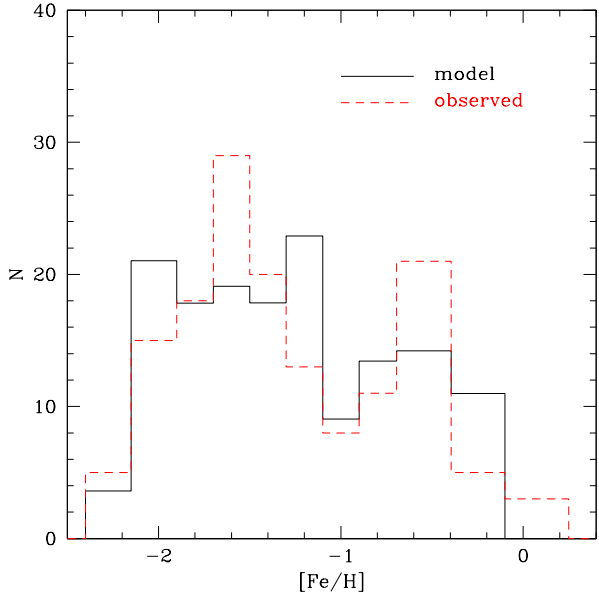}}
\caption{Metallicity distribution at $z=0$ in the best-fit model with
  an alternative prescription for $M_*$ as a function of $M_h$, given
  by equation (\ref{eq:fstar1}) ({\it solid histogram}), compared to
  the observed distribution of Galactic globular clusters ({\it dashed
  histogram}).}
\vspace{0.3cm}
  \label{fig:gc_metal2}
\end{figure}

The stellar fraction that we adopted from \citet{woo_etal08} is well
motivated by Milky Way dwarf galaxies at the present epoch, but the
abundance-matching models such as the one by \citet{conroy_wechsler09}
predict a steeper dependence on halo mass in the range $10^8\, \Msun <
M_* < 10^{10}\, \Msun$.  Additionally, the redshift dependence of this
relation is uncertain, and recent observational surveys
\citep{borch_etal06, bell_etal07, dahlen_etal07} have advocated a
slower evolution than $(1+z)^{-2}$ adopted in our model.  To
accommodate this uncertainty, we re-ran the entire parameter grid with
equation (\ref{eq:fstar1}) instead of equation (\ref{eq:fstar}).  The
corresponding contour plot is shown in
Figure~\ref{fig:p2p3contour_altmstar} and the best-fit metallicity
distribution is shown in Figure~\ref{fig:gc_metal2}.  This best fit is
capable of reproducing the observed metallicity ($P_{KS,Z} = 49\%$)
and mass distributions ($P_{KS,M} = 9.5\%$), similar to our fiducial
model.  The acceptable range of the parameter space is narrower and
shifted towards higher values of $p_2$, as the steeper mass slope
otherwise prevents low-mass halos from forming a sufficient number of
clusters.  Nevertheless, this alternative prescription still leads to
a significant chance of a bimodal metallicity distribution.

Current observational constraints on the gas fraction at high redshift
are even more uncertain.  We adjusted the fit by altering the scale
mass in equation (\ref{eqn:scalemass}).  As alternatives to the
fiducial model, we considered a power law of redshift, $M_s =
M_{s,0}\, (1+z)^2$, and an inverse time dependence, $M_s = M_{s,0}\,
(t/t_0)^{-1}$.  Both of these relations resulted in lower gas
fractions during the high-redshift epoch when most globular clusters
should form.  The gas fractions were too low for {\tt case-2}
formation for any reasonable choice of $p_5$.  More importantly, the
low gas masses did not allow even the most massive halos to form star
clusters until intermediate redshifts.  Therefore, none of these fits
is a viable alternative to the fiducial model.

The same problems manifested if we took the simplistic approach of
holding the gas fraction constant for all halo masses at all times.
This idea was initially considered to see if we could generate simple
results based only on halo merger histories without speculation on the
baryonic physics.  We quickly realized that this approach was not
going to work.  Setting $f_g$ too low effectively prevents cluster
formation at high redshift, when blue globular clusters are expected
to form, as most halos cannot build up sufficient mass to overcome the
minimum mass required to form a single massive star cluster (discussed
in Section \ref{sec:formation_rate}).  A constant gas fraction that is
too high, on the other hand, presents obvious unphysical predictions
at low redshift, and in particular would drastically over-predict the
number of young clusters, forcing us to arbitrarily truncate their
formation.

We also considered an alternative parametrization of the gas fraction,
suggested by \citet{stewart_etal09}.  They took the same observational
constraints as us, but fitted them as
\begin{equation}
  \frac{M_g}{M_*} = 0.04 \, 
      \left({M_* \over 4.5 \times 10^{11}\, \Msun}\right)^{-0.59
(1+z)^{0.45}}.
\end{equation}
This formula predicts so much cold gas at high redshift that many
low-mass halos would be able to form clusters via {\tt case-2} channel
for any $p_5<1$.  If we completely disable {\tt case-2} formation and
use the above prescription for the gas mass, we find many model
realizations consistent with the observed metallicity distribution.
This prescription differs from our fiducial choice in that it produces
considerably more young clusters and achieves less clear metallicity
bimodality.  The maximum value of the likelihood function attainable
with this prescription is approximately half of the value for the
fiducial prescription.  Nevertheless, it could still produce
acceptable globular cluster results.

In addition to changing the formulation of the fits, we investigated
the effect of adding a random Gaussian dispersion with standard
deviation $\sigma_{\rm fits}$ to the right hand sides of equations
(\ref{eq:charmass}), (\ref{eqn:mgms}), and (\ref{eq:fstar}) to reflect
their intrinsic scatter as well as observational
uncertainty. Different random values for the three scatters are
generated for each halo at each timestep, but we always force the
condition (\ref{eq:coldbar}) on the total baryon content.  For
simplicity, we used the same magnitude of $\sigma_{\rm fits}$ for the
scatter added to all three equations simultaneously.  We re-ran the
parameter search grid using $\sigma_{\rm fits} = 0.1, 0.2$, and 0.3
dex.  For each value of $\sigma_{\rm fits}$, we were still able to
find models with high values of $P_{KS,Z}$ and overall likelihood
statistic, although these values decline with the increasing amount of
scatter.  The metallicity probability varies from 49\% to 16\% to 6\%,
for $\sigma_{\rm fits} = 0.1, 0.2, 0.3$ dex respectively.

As an alternative to scatter in the cutoff mass
(eq.~\ref{eq:charmass}) with a fixed functional form for the gas
fraction (eq.~\ref{eq:fin}), we tried adding scatter to equation
(\ref{eq:fin}) while keeping fixed equation (\ref{eq:charmass}).
Adding scatter to $f_{\rm in}$ allows the gas fraction to exceed the
threshold $p_5$ much more easily and produce too many {\tt case-2} clusters.
To avoid unphysical results, we analyze only results for {\tt case-1}
formation.  In this case we find the best-fit models with $P_{KS,Z} =
47\%, 13\%$, and 5\%, for $\sigma_{\rm fits} = 0.1, 0.2, 0.3$ dex
respectively.  These models are still consistent with the observed
Galactic distribution.

The addition of scatter as described above has two systematic effects
on any individual realization of the metallicity distribution: the
high metallicity tail is extended even further and the height of the
blue peak is damped relative to the case of no scatter.  The former
effect is due to the possibility of drawing higher values of $M_*$ and
hence higher $\feh$.  The latter effect arises from the enforcement of
equation (\ref{eq:coldbar}), which prevents gas-rich halos at
high-redshift from gaining any extra gas from the positive scatter in
equation (\ref{eqn:mgms}); on the other hand, negative scatter can
prevent some of these halos from being eligible for {\tt case-2}
formation.  Accordingly, the best distributions with higher values of
$\sigma_{\rm fits}$ were found for models with low values of $p_5$.
We note that the Dip probability in most realizations is not strongly
affected by the new scatter, implying that the actual smearing of the
peaks is not significant and bimodality is preserved.

\subsection{Alternative Dynamical Disruption}
  \label{sec:altdyn}

In Section~\ref{sec:dyn} we noted that the expression for the
evaporation rate (eq.~[\ref{eq:nuev}]) contains some inherent
parameters.  Here we explore alternative disruption models with
different values of $\xi_e$ and $\delta$ within the fiducial formation
prescription.

The effect of decreasing $\xi_e$ is simply to reduce the number of
clusters that are completely disrupted by $z=0$.  In the fiducial
model with $\xi_e = 0.033$, about 60\% of the original sample is
disrupted.  (Note that this implies that roughly $5 \times 10^7\,
\Msun$ worth of stars in the Galactic stellar halo could be remnants
of the disrupted clusters.)  With the factor of $\xi_e = 0.02$, only
$\sim 30\%$ are disrupted.  With the factor of $\xi_e = 0.01$, almost
all clusters survive.

The effect of decreasing $\delta$ and $\delta_0$ is to shift the peak
of the mass function to a lower mass.

We repeated the grid parameter search for the best metallicity
distribution for two alternative prescriptions, one with $\xi_e =
0.02$, $\delta = \delta_0 = 1/3$, and the other with $\xi_e = 0.033$,
$\delta = \delta_0 = 1/9$.  We found that in both cases our model
could produce an observationally-consistent metallicity distribution.
However, lowering either $\xi_e$ or $\delta$ significantly alters the
mass function away from the data, by allowing too many low-mass
clusters to survive.  Raising $\xi_e$ and $\delta$ may improve the
mass function, but steers away from recent constraints on the two
parameters \citep{baumgardt_makino03, gieles_baumgardt08}.  Therefore,
we ultimately conclude that our fiducial prescription ($\xi_e =
0.033$, $\delta = \delta_0 = 1/3$) works best.

For illustration, we list below the properties of the best models in
the two alternative prescriptions.  For $\xi_e = 0.02$, $\delta =
\delta_0 = 1/3$, we find a peak model that has a metallicity
distribution with $P_{KS,Z} = 69\%$ and mass distribution with
$P_{KS,M} = 0.02\%$.  The parameters of this model are $p_2 = 4.4$,
$p_3 = 1.4$, $p_4 = 0$, $p_5 = 0.99$.

\begin{figure}[t]
\centerline{\epsfxsize3.3truein \epsffile{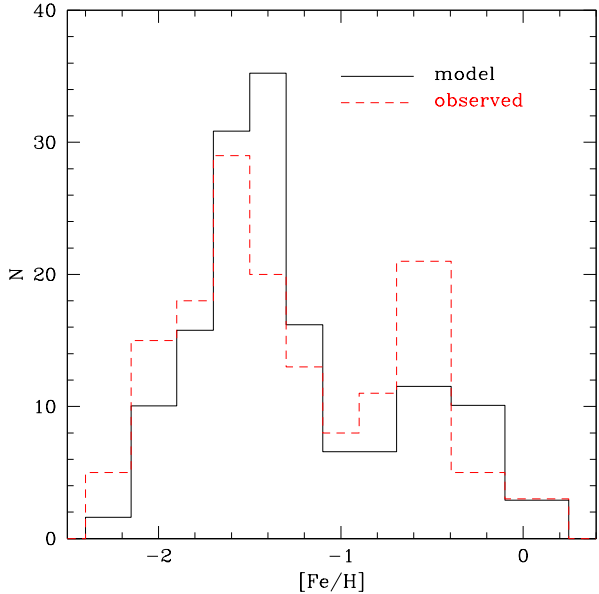}}
\caption{Metallicity distribution at $z=0$ in the best-fit model with
  an alternative dynamical disruption prescription,
  $\delta=\delta_0=1/9$ ({\it solid histogram}), compared to the
  observed distribution of Galactic globular clusters ({\it dashed
  histogram}).}
\vspace{0.3cm}
  \label{fig:gc_metal3}
\end{figure}

For $\xi_e = 0.033$, $\delta = \delta_0 = 1/9$, we find a peak model
that has a metallicity distribution with $P_{KS,Z} = 19\%$ and mass
distribution with $P_{KS,M} = 12\%$.  The parameters of this model are
$p_2 = 4$, $p_3 = 1.4$, $p_4 = 0$, $p_5 = 0.97$.  Its metallicity
distribution is shown in Figure~\ref{fig:gc_metal3}.  The
overabundance of metal-poor clusters is clear.

The latter alternative prescription ($\delta = 1/9$) predicts the
disruption time to scale with cluster mass in a manner nearly
identical to \citet{gieles_baumgardt08}, if we take that all model
clusters have a mean galactocentric frequency $\omega =
(2.4\times 10^7\, \mathrm{yr})^{-1}$.  Based on the orbit calculations
discussed in the next section, we find a similar median value of
$\omega$ for the sample of model clusters in the fiducial model.  We
also find no correlation between $\omega$ and cluster mass, which
confirms our assertion in Section \ref{sec:dyn} that in such a model
the disruption time of the average cluster would scale with mass as
$\nu_{\rm ev}^{-1} \propto M^{2/3}$.

\section{Spatial Distribution}
  \label{sec:spatial}

Using the same $N$-body simulation as in this paper,
\citet{prieto_gnedin08} investigated the present spatial distribution
of model clusters that formed in high-redshift (metal-poor) galactic
systems.  They calculated the orbits of clusters from the time when
their host galaxies accreted onto the main galaxy and identified three
distinct populations.  {\it Disk clusters} formed in the most massive
halo that eventually hosts the present Galactic disk.  These clusters,
found within the inner 10 kpc, are scattered into eccentric orbits by
the perturbations from accreted galactic satellites.  {\it Inner halo
clusters}, found between 10 and 60 kpc, came from the now-disrupted
satellite galaxies.  Their orbits are inclined with respect to the
Galactic disk and are fairly isotropic.  {\it Outer halo clusters},
beyond 60 kpc from the center, are either still associated with the
surviving satellite galaxies, or were scattered away from their hosts
during close encounters with other satellites and consequently appear
isolated.

The azimuthally-averaged space density of metal-poor globular clusters
is consistent with a power law, $n(r)\propto r^{-\gamma}$, with the
slope $\gamma \approx 2.7$.  Since all of the distant clusters
originate in progenitor galaxies and share similar orbits with their
hosts, the distribution of the clusters is almost identical to that of
the surviving satellite halos.  This power law is similar to the
observed slope of the metal-poor globular clusters in
the Galaxy.  However, the model clusters have a more extended spatial
distribution (larger median distance) than observed.  In the model it
is largely determined by the orbits of the progenitor galaxies and the
epoch of formation.  \citet{moore_etal06} showed that the
early-forming halos are more spatially concentrated and in order to
match the Galactic distribution, globular clusters would need to form
at $z \sim 12$.  However, such an early formation may be inconsistent
with the requirement of high mass and density of the parent molecular
clouds.

In this work, we have retraced some of these steps to attempt to
reproduce the spatial distribution of the whole Galactic globular
cluster system.

The clusters that formed in the disk of the main halo are assigned
radial positions according to the exponential profile, $dN/dR \propto
R \, e^{-R/R_d}, $ with the observed scale length of the Galactic
disk, $R_d = 3$ kpc.  The azimuthal angles are assigned randomly.  The
vertical position in the disk is also assigned randomly, with the
scale-height of one fifth of the scale-length.  The clusters are
limited to the radial range $0.6 < R < 10$ kpc, where the observed
disk globular clusters are located.  The distances are also given a
random Gaussian scatter of 10\% to replicate observational distance
uncertainties.

Clusters that formed in satellite halos that survived until $z=0$ are
assigned the present position of the host, with a small displacement
analogous to the distribution in the main disk.  Clusters that formed
in subhalos that did not survive until $z=0$ are initially assigned
the last known position and velocity of the host in the simulation,
with the same displacement as above.  We then follow the orbits of
these stray clusters until $z=0$ using a leap-frog integration scheme
with fixed time step.

The orbit integration follows \citet{prieto_gnedin08}.  The main halo
and the satellite halos contribute their NFW potentials, while the
disks within the halos contribute the Miyamoto-Nagai potentials with
the total mass of gas and stars computed from equations
(\ref{eqn:mgms}) and (\ref{eq:fstar}).  The total gravitational
potential is computed by linearly interpolating the masses of halos
and subhalos between the simulation snapshots.  Positions of subhalos
at each timestep are computed with cubic splines between the
snapshots.  We also include the acceleration on the clusters that
results from the use of the splines, as described in
\citet{prieto_gnedin08}.  Cosmological dark energy contributes an
additional component to the acceleration in physical coordinates:
${\bf a}_\Lambda = \Omega_\Lambda \, H_{0}^2 \, {\bf r}$.

Just as in the previous study, we find a more extended spatial
distribution of the globular cluster system than that observed in the
Galaxy.  Clusters that formed in surviving satellites (about 24\% of
the sample in the fiducial model) are the most distant from the
center, as forced by the location of the satellites.  The orbit
integration for the clusters formed in disrupted satellites (about
52\% of the sample) shows that these clusters also do not migrate in
$r$ far from the last known position of their host.  Such coupling to
the dark matter halos is the main reason for the overextended cluster
system.

Clusters that formed in the disk of the main halo (the remaining 24\%
of the sample) most closely resemble the spatial properties of the
Galactic clusters.  They are confined to the inner 10 kpc and would be
referred to as the bulge or disk clusters.  However, this group should
contain more than 50\% of the sample to be consistent with
observations.  Recent paper by \citet{griffen_etal10} similarly
investigated the formation of red clusters by major mergers in the
Aquarius simulation and concluded that such clusters must have formed
in the central disk.

Note that our orbit calculations, as well as those by
\citet{griffen_etal10}, use the gravitational potential derived in
collisionless cosmological simulations.  Stars and cold gas would
deepen the gravitational potential in the inner regions of the main
halo and bring the satellites closer to the center.  Dense stellar
nuclei of the satellites should also survive against tidal disruption
longer than pure dark matter halos.  Hydrodynamic simulations of
\citet{naab_etal09} show that the combined effect of baryons may be to
deposit half of stellar remnants of the disrupted satellites,
including their globular clusters, within 10 kpc of the center.  This
would effectively reconcile the predicted cluster distribution with
the Galactic sample, since over 50\% of our clusters formed in
disrupted satellites. More detailed hydrodynamic simulations of galaxy
formation are needed to verify either hypothesis.
 
An observational test of the cluster orbits would be possible when
proper motions are measured for a large fraction of the Galactic
clusters.  Such measurements could be achieved with the planned
SIM-Lite space observatory.

\section{Globular Cluster Colors}
\label{sec:color}
We attempted another direct comparison of the model predictions with
the observed sample, by constructing single stellar population models
using \citet{bruzual_charlot03} code GALAXEV.  As input for GALAXEV,
we used the age and metallicity for each globular cluster created with
the fiducial model.

The distribution of the model $B-V$ color was considerably less
bimodal than the metallicity distribution discussed in previous
sections.  The main cause of the smearing of the two peaks appears to
be the younger age of metal-rich clusters predicted by the model.  The
metal-poor clusters constitute a clearly defined peak at $B-V = 0.67$,
which corresponds well to the blue peak of the Galactic sample.  But
the metal-rich clusters, which are expected to make up the red peak,
have a mean $B-V$ color of 0.77, while the observed red mean is close
to 0.85.  The standard deviation of the red model clusters is 0.08,
implying that the result is consistent within one sigma of the
observed, but the bimodality of the distribution is not evident to the
eye.

To test the hypothesis that the smearing of the color peaks compared
was due to the relative age of the populations, we ran the population
synthesis models again, this time using a constant age of 12.1 Gyr for
all clusters.  The resulting distribution indeed appeared to
constitute two peaks, with a blue peak at a mean of $B-V = 0.67$ and a
red peak at $B-V = 0.84$, with a clearly defined gap between them. It
should be noted that a known discrepancy exists between the $B-V$
colors predicted by all major population synthesis codes and those of
observed globular clusters \citep{conroy_gunn09}. All models predict
colors that are too blue at high metallicity, which would directly
play into smearing bimodality in our result.

In addition to the colors, we examined the cluster luminosities
(absolute $V$-band magnitudes) calculated by GALAXEV.  These allow a
more direct comparison with the observations than the mass function
presented in Section \ref{sec:dyn}, for which we were required to
assume a constant mass-to-light ratio for all observed clusters.  This
constant $M/L_V$ ratio has been a traditional approach, but has come
under recent scrutiny by \citet{kruijssen08} who argued that $M/L_V$
may vary as a function of cluster mass.  However, the distribution of
$V$-band magnitudes for our fiducial model has a KS probability of
4.3\%, which is not a significant departure from the 7.4\% for the
mass function.  When we tried the same exercise as above by setting
the ages of all model clusters to 12.1 Gyr, the KS probability jumped
to 25\%.  This improvement likely happened because we converted the
magnitudes of observed clusters into masses using $M/L_V = 3$, while
GALAXEV typically predicted $M/L_V < 3$ for the 12.1 Gyr
isochrones. This brought the mean luminosity of model clusters closer
to the observed value than the average mass of model clusters was to
its observed counterpart.

Even though these population synthesis results are interesting, we
believe that the mass and metallicity distributions presented in
previous sections are more reliable.  Population synthesis modeling
adds an extra layer of empirical uncertainty to our results, as the
specific nature of horizontal-branch evolution remains an issue that
has not been completely resolved.

\section{Summary and Implications for Galaxy Formation Models}
  \label{sec:discussion}

We have presented a model for the origin of the metallicity
distribution of globular clusters.  In our scenario, bimodality
results from the combination of the history of galaxy assembly (rate
of mergers) and the amount of cold gas in protogalactic systems.
Early mergers are frequent but involve relatively low-mass
protogalaxies, which produce preferentially blue clusters.  Late
mergers are infrequent but typically involve more massive galaxies.
As the number of clusters formed in each merger increases with the
progenitor mass, just a few late massive mergers can produce a
significant number of red clusters.  The concurrent growth of the
average metallicity of galaxies between the late mergers leads to an
apparent ``gap'' between the red and blue clusters.

The peak metallicities of the red and blue populations are remarkably
robust to variations of the model parameters.  The peaks encode the
mass-metallicity relation in galaxies and do not depend strongly on
the rate or timing of cluster formation.  The exact definition of a
major merger is also not important for our result, as long as the
merger mass ratio is at least 1:5.

Our conclusions on the origin of metallicity bimodality are not
significantly affected by the large uncertainties in our knowledge of
the stellar mass and cold gas mass in high-redshift galaxies.  We
considered alternative prescriptions for the stellar fraction,
gas-to-stars ratio, and even dynamical disruption, but in all cases
found a metallicity distribution consistent with the observations.
Such robustness indicates that most external factors are not as
important as the internal mass-metallicity relation in host galaxies.

We find that dynamical disruption over the cosmic history naturally
converts an initial power-law cluster mass function into an observed
log-normal distribution.  A continuous formation of clusters in the
first several Gyr help to replenish the depleted low-mass end.
Dynamical disruption also helps establish metallicity bimodality by
preferentially depleting old clusters in the metal-poor peak.

Our prescription links cluster metallicity to the average galaxy
metallicity in a one-to-one relation, albeit with random scatter.
Since the average galaxy metallicity grows monotonically with time,
the cluster metallicity also grows with time.  Our model thus encodes
an age-metallicity relation, in the sense that metal-rich clusters are
somewhat younger than their metal-poor counterparts.  Observations of
the Galactic globular clusters indicate an age spread that ranges from
1 Gyr for the inner blue clusters to 2 Gyr for the inner red clusters
to 6 Gyr for the outer clusters, which is generally consistent with
the predicted spread.  However, the model may be marginally
inconsistent with the observation that some of the metal-rich clusters
appear as old as the metal-poor ones.  Note that our model is still
simplistic and does not include metallicity gradients within
protogalaxies, which may dilute the predicted age-metallicity
relation.

Our model demonstrates that star cluster formation during gas-rich
mergers of protogalactic systems is a single mechanism that
successfully reproduces many observed properties of the Galactic
globular clusters.  It may avoid the need for two separate formation
mechanisms for the red and blue clusters invoked in the model of
\citet{beasley_etal02}.  Their model relied on constant cluster
formation efficiency relative to the field stars, but required
different efficiencies for the two modes.  While the red clusters in
their models continued forming through galaxy mergers, the blue
clusters needed to be arbitrarily shut off at $z=5$.  This was the
main cause of the bimodal metallicity distribution in their model, as
the blue clusters did not have much overlap with the red clusters that
formed in major mergers involving metal-enriched gas considerably
after $z=5$.  The \citet{beasley_etal02} model also neglected the
effects of the dynamical evolution that shaped the present cluster
distribution.  In contrast, in our model some old blue clusters are
disrupted and some are unable to form at recent times because the
protogalaxies are gas-poor.  Another difference is that in our model
major mergers contribute both red and blue clusters, while in Beasley
et al. they contribute only red clusters.  We also find that globular
clusters form significantly earlier than the bulk of field stars and
therefore the two cannot be linked by a constant formation efficiency
at all times (see Fig.~\ref{fig:MGCMstar}).

We have compared the metallicity distribution of globular clusters to
the mass-weighted metallicity distributions of other stellar
populations as predicted by our scaling relations given in Section 2.
We find that galaxy field stars overall have a single-peaked
distribution with a mean of $\feh \approx 0$, a metal poor tail, and
no stars with $\feh > 0.4$.  This is consistent with our current
understanding of the metallicity of stars in the Galactic disk.  The
stars in surviving satellites, which correspond to Milky Way dwarf
galaxies, also appear to have a single-peaked distribution with a mean
metallicity $\feh \approx -1$.  Only the globular cluster system
display a bimodal metallicity distribution.

We derived some simple scaling relations for the overall efficiency of
globular cluster formation.  We adopted the cluster formation rate in
gas-rich, high-redshift merger events (eq.~\ref{eqn:massGC}) that
scales with the host system mass as $M_{GC} \sim 10^{-4}\, M_g/f_b
\sim 10^{-4}\, M_h$.  We have later learned of a similar empirical
relation for all types of massive galaxies, derived independently by
\citet{spitler_forbes09} and \citet{georgiev_etal10}.  The outcome of
the model is a prediction that the fraction of galaxy stellar mass
locked in star clusters, $M_{GC}/M_*$, is of the order $10-20\%$ at
$z>3$ and then declines steadily with time to about 0.1\% at present.
The specific frequency parameter follows a similar decline with time
and reaches $N/(M_*/10^9 \, \Msun) \sim 1$ at the present.  These
efficiencies are in agreement with the compilations of
\citet{mclaughlin99}, \citet{rhode_etal05}, and \citet{peng_etal08}.
We also find that the globular cluster system overall is significantly
more metal-poor than the galactic spheroid, which is populated by
stars from the disrupted satellites.

Our scenario can be applied to other galactic environments, such as
those of elliptical galaxies which contain much larger samples of
globular clusters.  For example, \citet{peng_etal08} showed that the
fraction of red clusters increases from 10\% to 50\% with increasing
luminosity of elliptical galaxies in the Virgo cluster.  In our model,
globular cluster formation is entirely merger-driven.  We showed that
the Galactic sample may have arisen from early super-gas-rich low-mass
mergers and later metal-rich high-mass mergers.  Compared to the
Galaxy, giant ellipticals are expected to experience more high-mass
mergers which would contribute more prominently to the globular
cluster system.  As Figure~\ref{fig:gc_met_case} shows, such mergers
would produce comparable numbers of red and blue clusters
simultaneously.  Thus the fraction of red clusters should increase
with galaxy mass, reaching $\sim 50\%$ for giant ellipticals.  This
trend, observed by \citet{peng_etal08}, may be a natural outcome of
the hierarchical formation.

At the other end of the galactic spectrum, dwarf galaxies likely
lacked metal-rich mergers and produced only metal-poor blue clusters.
In particular, dE and dSph type dwarfs which are now deprived of cold
gas are not expected to contain any young and metal-rich clusters.
Some dIrr galaxies, such as the LMC, still possess considerable
amounts of cold gas and may produce younger clusters, although they
are still likely to have subsolar metallicity.  The variety of
globular cluster ages observed in the LMC indicates that it may have
had bursts of star formation throughout its cosmic evolution.

Our study places interesting constraints on galaxy formation models.
Within the framework of our model, acceptable mass and metallicity
distributions result only from a certain range of the parameters.  In
particular, the minimum ratio of masses of merging protogalaxies
strongly correlates with the cluster formation rate.  If the clusters
form very efficiently only a few massive mergers are needed; if the
clusters form inefficiently many mergers are needed, which requires a
lower merger threshold.  However, mass ratios of less than 0.2 are
disfavored in the model (see Fig.~\ref{fig:p2p3contour}).  Formation
of massive clusters in very gas-rich systems without detected mergers
(our {\tt case-2} scenario) improves the final mass function but is
not required for reproducing the metallicity distribution.  Thus,
globular cluster formation solely in major mergers is consistent with
the available observations.  Finally, our results rest on the derived
prescription for the cold gas fraction as a function of halo mass and
cosmic time.  This prescription (Fig.~\ref{fig:mgas}) can be tested by
future observations of high-redshift galaxies with JWST and by
detailed hydrodynamic simulations.

\acknowledgements 

We would like to thank Kerby Shedden for a discussion of statistical
methods regarding bimodality, Steve Zepf for a copy of the original
KMM code, Karl Gebhardt for a discussion of the Dip test, Martin
Maechler for an updated probability table of the Dip test, Evan Kirby
for a discussion of metallicities of the ultrafaint dwarfs, and Eric
Bell, Nick Gnedin, Andrey Kravtsov, and Lee Spitler for comments on
the manuscript.

\appendix
\section{Quantifying Bimodality}

Quantifying whether a distribution is better described by one or two
modes is still an unsolved problem in statistics.  While there are
algorithms that split the input distribution into two modes or assign
probabilities that a given data point belongs to either of the two
modes, there is no proper statistic that evaluates whether such a
split is preferred to a unimodal distribution.  In this Appendix, we
describe a popular KMM algorithm and our improvement of it, as well as
an independent test of bimodality based on the Dip statistic.

\subsection{GMM -- A better version of KMM}

\citet{ashman_etal94} popularized a mixture modeling code KMM for
detecting bimodality in astronomical applications.  This code has been
widely used for globular cluster studies and can be considered a
standard method in the field.  The KMM algorithm assumes that an input
sample is described by a sum of two Gaussian modes and calculates the
likelihood of a given data point belonging to either of the two modes.
It also calculates the likelihood ratio test (LRT) as an estimate of
the improvement in going from one Gaussian to two Gaussian
distributions.  However, the LRT obeys a standard $\chi^2$ statistic
only when the two modes have the same width (variance), which may not
be satisfied by real datasets.  Even though the probability of the LRT
can be estimated using bootstrap in principle, in practice the use of
the KMM code has been limited to common width modes (the so-called
``homoscedastic'' case).  \citet{brodie_strader06} and
\citet{waters_etal09} provide further discussion of KMM.

The KMM method belongs to a general class of algorithms of Gaussian
mixture modeling (GMM).  GMM methods maximize the likelihood of the
data set given all the fitted parameters, using the
expectation-maximization (EM) algorithm
\citep[e.g.,][]{numerical_recipes}.  A major simplification, which
allows one to derive explicit equations for the maximum likelihood
(ML) estimate of the parameters, is that each mode is described by a
Gaussian distribution.

For simplicity, and as appropriate for the metallicity distribution,
we consider a univariate input data set.  However, the algorithm is
fully scalable to multivariate distributions.  The likelihood function
of a univariate sample $x_n$ is
\begin{equation}
  {\cal L}_K = \prod_n \left( \sum_{k=1}^K \, p_k \, N(x_n |
\mu_k,\sigma_k) \right),
\end{equation}
where
\begin{equation}
  N(x | \mu,\sigma) = {1 \over (2\pi \sigma^2)^{1/2}}
      \exp{\left[-{(x-\mu)^2 \over 2\sigma^2}\right]}
\end{equation}
is the Gaussian density.  The modal fractions are normalized as
$\sum_k p_k =1$.  A unimodal distribution ($K=1$) has two independent
parameters ($\mu$ and $\sigma$), whereas a bimodal distribution has
five parameters ($p_1, \mu_1, \sigma_1, \mu_2, \sigma_2$), since
$p_2=1-p_1$.

The power of the GMM method lies in its ability to determine the ML
values of the parameters ($p_k, \mu_k, \sigma_k$).  The disadvantage
is that the method will always split the data set into the specified
number of modes, $K$.  In order to detect bimodality it is extremely
important to be able to judge whether the bimodal fit is an
improvement over the unimodal fit.  For this purpose the KMM code uses
the LRT test, which appears to be an approximation derived by
\citet{wolfe71} for the homoscedastic case ($\sigma_1 = \sigma_2$).
Define the ratio of the maximum likelihoods as $\lambda \equiv {\cal
L}_{1,\rm max}/{\cal L}_{2,\rm max}$.  According to numerical Monte
Carlo studies of \citet{wolfe71}, the statistic $-2 \ln{\lambda}$
approximately obeys the $\chi^2$ distribution with a number of degrees
of freedom equal to ``twice the difference between the number of
parameters of the two models under comparison, not including the
mixing proportions'' \citep{mclachlan87}.  This is the $\chi_2^2$
distribution in our case.  However, the statistic does not apply in
the heteroscedastic case $\sigma_1 \neq \sigma_2$ (it would have been
$\chi_4^2$).  Note that this unusual number of degrees of freedom was
found as an empirical approximation.  Unfortunately, no exact
estimation exists for the goodness of modal split.

Several variations of the method have been suggested in the
literature.  \citet{mclachlan87} proposed a {\it parametric bootstrap}
to test for the number of components.  In this method, a test sample
is drawn randomly from a unimodal Gaussian distribution with the
parameters \{$\mu$, $\sigma$\} best-fitting the input sample.  The
number of objects in the test sample is taken the same as in the input
sample.  The bimodal split is calculated for this test sample using
the EM algorithm and the likelihood ratio $\lambda_{\rm boot}$ is
saved.  Repeating the bootstrap a large number of times, we obtain the
probability of randomly drawing the ratio as large as that observed in
the input sample, $\lambda_{\rm obs}$.  If the probability is below a
few percent, we reject the null hypothesis that the input sample
belongs to a unimodal Gaussian.

However, the parametric bootstrap is not a perfect solution.  In the
limit of a large number of objects in the input sample, the likelihood
function is very sensitive to outliers far from the center of the
distribution.  Simple measurements errors in the wings of the Gaussian
function may cause a unimodal distribution to be rejected, even if
it is correct.  In other words, GMM is more a test of Gaussianity
than of unimodality \citep[see][for more discussion]{muthen03,
bauer07}.

\citet{lo_etal01} proposed a modified LRT method to test for the true
number of components of a Gaussian mixture.  The modified statistic
must be evaluated numerically, but still does not address the problem
with Gaussian wings.  Subsequently, \citet{lo08} suggested to use the
standard LRT with the parametric bootstrap to test for heteroscedastic
split, and also suggested restricting the ratio of the standard
deviations of the two modes to be not less than 0.25, to avoid
numerical artifacts.  Such a method was recently implemented in
globular cluster studies by \citet{waters_etal09}.

The sensitivity of LRT to the assumption of Gaussian distribution
calls for additional, independent tests of bimodality.  A useful and
intuitive statistic is the separation of the means relative to their
widths:
\begin{equation}
  D \equiv {|\mu_1 - \mu_2| \over \left[ (\sigma_1^2 + \sigma_2^2)/2
\right]^{1/2}}.
  \label{eq:dpeak}
\end{equation}
We use the factor $\sqrt{2}$ for consistency with the definition in
\citet{ashman_etal94}, who noted that $D>2$ is required for a clean
separation between the modes.  If the GMM method detects two modes but
they are not separated enough ($D<2$), then such a split is not
meaningful.  The power of GMM in this case is counterproductive.  A
histogram of such a distribution would show no more than two little
bumps, which would not be recognized as distinct populations.

Another simple statistic is the kurtosis of the input distribution.  A
positive kurtosis corresponds to a sharply peaked distribution, such
as the Eiffel Tower.  A negative kurtosis corresponds to a flattened
distribution, such as a top hat.  A sum of two populations, not
necessarily Gaussians, is broader than one population and therefore
has a significantly negative kurtosis.  However, $kurt < 0$ is a
necessary but not sufficient condition of bimodality.  A broad
unimodal distribution, such as an actual top hat, also has negative
kurtosis.  Therefore, $kurt < 0$ is only useful as an additional check
to support the results of LRT and the $D$-value.

In order to provide a more robust measure of the modal split, we have
revised and implemented the GMM algorithm independently of the KMM
code.  We begin with a single run of the EM algorithm to calculate the
means and standard deviations assuming a heteroscedastic bimodal
distribution.  Then we repeat the estimation assuming a unimodal
Gaussian case.  We take the ratio of the likelihoods $\lambda$, the
separation $D$, and the kurtosis as the three statistics of
interest.  We then estimate the error distribution for the modal
parameters using non-parametric bootstrap (drawing from the input
sample with repetitions) of 100 realizations.  We also run the
parametric bootstrap to assess the confidence level at which a
unimodal distribution can be rejected based on each of the three
statistics.  Its practical application for an input sample of more
than 100 objects is limited to about 1000 bootstrap realizations,
limiting the confidence level to $\sim 10^{-3}$.  A sufficiently low
probability of each statistic means a unimodal distribution can be
rejected in favor of a bimodal distribution.  The code also calculates
the probability of each data point belonging to either mode.

A sum of two Gaussians with the same variance can sometimes be
preferred to the case of different variances, because of the one fewer
degree
of freedom.  The choice between the homoscedastic and heteroscedastic
cases can be similarly made using LRT, but we feel that it is less
important than choosing between a bimodal and unimodal distributions.
For comparison with the KMM code, we calculate the homoscedastic split
and its approximate probability using the $\chi_2^2$ statistic.  We
also calculate an alternative split into two Gaussian modes with the
same mean but different variance.  This is a more extended
distribution than a single Gaussian, which may be a better fit for a
unimodal but non-Gaussian sample.

The steps of our algorithm are summarized below:

1. Calculate $\{\mu, \sigma\}$ and $\{p_1, \mu_1, \sigma_1, \mu_2,
\sigma_2\}$ using a single EM run.

2. Form three statistics: $\lambda$, $D$, and $kurt$.

3. Run non-parametric bootstrap to estimate the errors:
   $\Delta\mu_k$, $\Delta\sigma_k$, $\Delta p_1$, and $\Delta D$.

4. Run parametric bootstrap to estimate the probability of a unimodal
   distribution, according to $\lambda$, $D$, and $kurt$.

As a first test of the algorithm, we verified that it reproduces
exactly the test output of the KMM code, given the test input provided
with the code.

We also made two random realizations of a unimodal Gaussian
distribution, $N(0,1)$, with 150 objects and 1500 objects,
respectively.  The smaller sample has the mean and standard deviation
of $\mu = -0.088$ and $\sigma = 0.987$, within the intended target
given the sample size.  Indeed, a non-parametric bootstrap gives
$\Delta\mu = 0.084$, $\Delta\sigma = 0.063$.  The kurtosis of the
input sample is $kurt=0.104$.  A heteroscedastic split gives two
peaks, by construction, with $\mu_1 = -0.483$, $\sigma_1 = 1.206$ and
$\mu_2 = -0.032$, $\sigma_2 = 0.938$.  However, the split is not
statistically significant.  The likelihood is improved only by $-2
\ln{\lambda} = 0.26$ relative to the unimodal case, which gives the
probability better than 99\% that the input sample is unimodal.  The
parametric bootstrap gives a similar probability of 96\%.  The
separation of the peaks also leads to the same conclusion: $D = 0.42
\pm 1.46$.  The parametric bootstrap probability of drawing such value
of $D$ randomly from a unimodal distribution is 87\%.  The probability
of drawing the measured kurtosis is 74\%.  Thus, all three statistics
show correctly that the input distribution is not bimodal.  The larger
test sample has smaller parameter errors, as expected, but similar
significance levels from the parametric bootstrap.

We then apply the GMM algorithm to the sample of observed
metallicities of the Galactic globular clusters.  A unimodal fit gives
$\mu = -1.298 \pm 0.049$ and $\sigma = 0.562 \pm 0.028$, where the
errors are calculated with the non-parametric bootstrap.  A
heteroscedastic split gives $\mu_1 = -1.608 \pm 0.064$, $\sigma_1 =
0.317 \pm 0.051$ and $\mu_2 = -0.583 \pm 0.074$, $\sigma_2 = 0.281 \pm
0.075$.  Of the total number of 148 clusters, 103 (or 70\%) are in the
metal-poor group and 45 (or 30\%) are in the metal-rich group.  A
homoscedastic split gives $\mu_1 = -1.620 \pm 0.037$, $\mu_2 = -0.608
\pm 0.055$, and $\sigma_1 = \sigma_2 = 0.303 \pm 0.026$.  In this
case, there are 101 metal-poor clusters and 47 metal-rich clusters.
In either case, the likelihood improvement in the 1000 parametric
bootstrap realizations is never as high as observed, $-2 \ln{\lambda}
= 27.5$.  That is, a unimodal distribution is rejected at a confidence
level better than 0.1\%.  The separation of the peaks is also very
clear, $D = 3.42 \pm 0.47$.  The observed cluster distribution is
indeed bimodal!

When we apply the GMM algorithm to our model sample, we also find
strong bimodality.  In order to keep the sample size similar to the
data, we test separately each of the 11 random model realizations of
each host halo.  The resulting parameters form a set from which we
calculate the average parameters.  To estimate both the statistical
and systematic errors of each parameter, we sum in quadrature its mean
standard deviation in the set and the scatter among the realizations.
A heteroscedastic split gives $\mu_1 = -1.543 \pm 0.045$, $\sigma_1 =
0.318 \pm 0.043$ and $\mu_2 = -0.576 \pm 0.077$, $\sigma_2 = 0.237 \pm
0.026$.  The break-down is 66\% ($\pm 4\%$) of the clusters in the
metal-poor group and 34\% in the metal-rich group.  The average
separation of the peaks is very significant, $D = 3.44 \pm 0.29$.
None of the 1000 parametric bootstrap sets of a unimodal distribution
returned any of the three statistics ($\lambda$, $D$, $kurt$) as high
as those found the model.  Again, a unimodal distribution is rejected
at a confidence level better than 0.1\%.  Thus, our fiducial
metallicity distribution is bimodal as well.

A different methodology to the LRT in the Bayesian approach is the
Odds Ratio, or the Bayes factor described in \citet{liddle09}.  The
Odds Ratio is the ratio of the integrals of the likelihood function
${\cal L}$ over all possible ranges of the model parameters with the
corresponding normalized distribution functions of the parameters.  In
the case of one-Gaussian vs. two-Gaussians, it is $\int {\cal L}_2
P(p_1) P(\mu_1) P(\sigma_1) P(\mu_2) P(\sigma_2) dp_1 d\mu_1 d\sigma_1
d\mu_2 d\sigma_2 / \int {\cal L}_1 P(\mu) P(\sigma) d\mu d\sigma$.
The Odds Ratio is dimensionless and is greater than 1 if the modal
split indeed significantly improves the likelihood ${\cal L}_2$ over
${\cal L}_1$.  On the other hand, if the improvement in the likelihood
is small, it can be washed out by integration over the additional
parameters and the Odds Ratio becomes less than 1.  We have
experimented with using the Odds Ratio as an objective criterion for
the goodness of modal split but found that its application depends
sensitively on the adopted range of the parameters and their (unknown)
distribution functions.  Therefore, we have not included it in our GMM
code.

\subsection{Dip test}

A completely independent test of unimodality was proposed by
\citet{hartigan_hartigan85}.  It was first used for globular cluster
studies by \citet{gebhardt_kissler-patig99}.  The Dip test is based on
the cumulative distribution of the input sample.  The Dip statistic is
the maximum distance between the cumulative input distribution and the
best-fitting unimodal distribution.  In some sense, this test is
similar to KS test but the Dip test searches specifically for a flat
step in the cumulative distribution function, which corresponds to a
``dip'' in the histogram representation.  The probability of rejecting
a unimodal distribution is calculated empirically and tabulated as a
function of sample size.  We obtained an updated table of the
probabilities calculated recently by Martin Maechler
(www.cran.r-project.org/web/packages/diptest).

We have added a driver routine to the original Fortran code of
\citet{hartigan_hartigan85}.  Our code interpolates the probability
table for any input sample size up to 5000 objects.  Looking just at
the significance levels, the Dip test appears less powerful than GMM.
The Dip probability of the observed Galactic sample being bimodal is
90\%, whereas the LRT probability is 99.998\% and the parametric
bootstrap probability is 99.9\%.  However, the Dip test has the
benefit of being insensitive to the assumption of Gaussianity and is
therefore a true test of modality.  It is also much faster to run than
the GMM code.  We use the Dip probability for assessing bimodality of
our model realizations in Section~\ref{sec:bimodality}.

Note that we searched for a statistical method that evaluates
bimodality using full input information, without having to bin the
data.  Another method, RMIX, based on a histogram of the input sample
was recently suggested by \citet{wehner_etal08}.

Source codes of the GMM and Dip tests are available from one of the
authors (OG) upon request.

\makeatletter\@chicagotrue\makeatother

\bibliography{gc}

\end{document}